\documentclass[12pt]{article}

\RequirePackage[OT1]{fontenc}
\RequirePackage{amsthm,amssymb,amsfonts,graphicx,subfigure,caption,bm,fullpage,wrapfig,setspace,color}
\RequirePackage[colorlinks,citecolor=blue,urlcolor=blue]{hyperref}
\usepackage{amsmath,amssymb,hyperref}

\numberwithin{equation}{section}
\theoremstyle{plain}
                          \newtheorem{thm}{Theorem}
                          
\theoremstyle{remark}         
\theoremstyle{definition} 

\newcommand\om{\Omega}
\newcommand\real{\mathbb{R}}

\newcommand\T{\mathcal{T}}
\renewcommand\S{\mathcal{S}}

\newcommand\bS{\bm{S}}

\newcommand\bx{\bm{x}}

\newcommand\bnu{\bm{\nu}}

\newcommand\bX{\bm{X}}

\newcommand\brho{\bm\rho}
\newcommand\balp{\bm\alpha}

\newcommand\bphi{\bm\phi}

\newcommand\trho{\tilde{\rho}}

\newcommand\tvarphi{\tilde{\varphi}}

\newcommand\tbrho{\tilde{\vect{\rho}}}

\newcommand\bi{\begin{itemize}}
\newcommand\ei{\end{itemize}}

\newcommand{\vect}[1]{\mbox{\boldmath $ #1$}}
\newcommand{\iid}{\stackrel{\mathrm{iid}}{\sim}}
\newcommand{\ind}{\stackrel{\mathrm{ind}}{\sim}}

\def\I{{\mathbf{1}}}

\def\woMR#1{\w@MR#1MR#1MR\relax}%
\def\w@MR#1MR#2MR#3\relax{#2}

\def\@MR#1 #2\relax#3{%
 \href{http://www.ams.org/mathscinet-getitem?mr=#1}%
 {\MRfixed{#3}}}%

\def\MRfixed{MR\woMR}%

\usepackage{natbib}

\usepackage{varioref}		
\labelformat{section}{Section~#1}
\labelformat{figure}{Figure~#1}
\labelformat{table}{Table~#1}	

\title{Analysis of distributional variation through multi-scale Beta-Binomial modeling}
\author{Li Ma and Jacopo Soriano}

\begin{document}
\maketitle
\doublespacing 

\begin{abstract}
Many statistical analyses involve the comparison of multiple data sets collected under different conditions in order to identify the difference in the underlying distributions. 
A common challenge in multi-sample comparison is the presence of various confounders, or extraneous causes other than the conditions of interest that also contribute to the difference across the distributions. They result in false findings, i.e., identified differences that are not replicable in follow-up investigations. We consider an ANOVA approach to addressing this issue in multi-sample comparison---by collecting replicate data sets under each condition, thereby allowing the identification of the interesting distributional variation from the extraneous ones. We introduce a multi-scale Bayesian hierarchical model for the analysis of distributional variation (ANDOVA) under this design, based on a collection of Beta-Binomial tests targeting variations of different scales at different locations across the sample space. Instead treating the tests independently, the model employs a graphical structure to introduce dependency among the individual tests thereby allowing borrowing of strength among them. We derive efficient inference recipe through a combination of numerical integration and message passing, and evaluate the ability of our method to effectively address ANDOVA through extensive simulation. We utilize our method to analyze a DNase-seq data set for identifying differences in transcriptional factor binding.
\end{abstract}


\section{Introduction}
\vspace{-1em}

An inference task that holds key to numerous applications is the comparison of several data sets to identify the difference in the underlying distributions. The most common setting is the $k$-sample problem, in which one compares i.i.d.\ data samples from $k$ unknown distributions 
supported on a common sample space $\om$.
The samples may be collected under $k$ different experimental settings, treatment statuses, or response labels, etc. The special case with $k=2$ forms the so-called two-sample problem, which has received tremendous attention and has been studied extensively from various perspectives throughout the past decades. See for example \cite{anderson:1962,schilling_1986,henze_1988,holmes_etal_2012,ma_wong_2011,chen_hanson_2012}.

In many practical applications of $k$-sample studies, however, the identified differences are often {\em not} replicable in follow-up investigations. 
One frequent cause of such false findings is confounding---the presence of extraneous sources of variation that causes cross-sample differences but are not controlled in the study. 
This is particularly prevalent when the interesting variation is of similar or smaller magnitude than that of the extraneous variability.
We shall provide one such example from genomics---namely the comparison of DNase-seq count profiles \citep{song:2010,degner:2012}---in our data analysis, in which the objective is to identify differences in DNase-seq count profiles that reveal differential transcriptional factor binding across multiple cell lines. 

One strategy to addressing confounding is through a study design that allows the interesting variations, which we shall refer to as the {\em intrinsic}  variation, to be identified from the uninteresting, {\em extraneous} ones. The simplest choice is the $k$-group design, analogous to the (one-way) ANOVA in classical experimental design. Under this design, $k$ {\em groups} of data sets are collected under different conditions, each group consisting of multiple replicate samples under the same condition, reflecting the extraneous variation. The rest of the work will concern the $k$-group design, and so we use ``intrinsic variation'' interchangeably with ``{\em cross-group} variation'', and ``extraneous variation'' interchangeably with ``{\em in-group} variation''.  

We focus on developing a nonparametric method for analyzing data collected this way. The inference objective is similar in spirit to the classical ANOVA. The key difference is that now each ``observation'' is a data sample from an unknown distribution, and the interest is in the variance components of the {\em distributions} themselves and not of the data they generate. 
We shall refer to this problem as the {\em analysis of distributional variation} (ANDOVA). 

Our main goal is to design an approach to ANDOVA flexible enough to accommodate various forms of distributional variation while achieving analytical simplicity and computational efficiency. In particular, the approach should be capable of incorporating the spatial heterogeneity (SH) of the underlying variation, a prevalent phenomenon in ANDOVA problems observed (to various extents) in virtually all modern applications.
Specifically, SH refers to the changing magnitude of variation, both intrinsic and extraneous, across the sample space. A common form of SH is the local nature of the cross-group variation. That is, often only a small portion of data or the sample space is affected by the different conditions in any scientifically significant way. On the other hand, 
because the extraneous sources of variation are typically a hodge-podge of confounders, their contribution to distributional variation typically affects most, if not all, of the sample space. But even then the extent of such contribution can vary substantially across the space.

With these considerations in mind, we adopt a general multi-resolution scanning strategy \citep{walther2010optimal,rufibach2010block,soriano&ma:2015}---to divide the sample space through windows at different locations and of different sizes and infer on each window for cross-group variation while adjusting for the in-group variation therein. In particular, we apply a hypothesis test on each window based on a Beta-Binomial (BB) model that characterizes the variation at a given location and scale.
We encapsulate the totality of all window-specific BB models into a joint multi-scale hierarchical model called the multi-scale Beta-Binomial (ms-BB) model.
We extend the model by adding a hyperprior in the form of a Markov tree \citep{crouse_etal_1998} to incorporate dependency across the window-specific BB models. This allows borrowing strength across windows, enhancing the ability to identify small variations that individual windows do not provide strong enough empirical evidence for reliable inference. We show that the joint posterior of this ``graphical ms-BB'' model can be computed using a combination of numerical integration and message passing \citep{wainwright:2015}, allowing Bayesian inference to proceed computationally efficiently.

In \ref{sec:method} we present our method for ANDOVA. We start in Section~\ref{sec:multiscale_decomp} with a brief review on the two building blocks for our method---the classical Beta-Binomial model and a multi-scale decomposition of probability distributions. In Section~\ref{sec:ms_ANDOVA} we use them to build the ms-BB model. In Section~\ref{sec:graphical_ms_ANDOVA} we extend the model to incorporate the graphical hyperprior. We then discuss guidelines on prior specification as well as decisions rules for identifying significant cross-group variations in Section~\ref{sec:prior_spec_multiplicity}. In \ref{sec:numerical_examples} we evaluate the performance of our method through simulations under various scenarios. In \ref{sec:dnase} we illustrate the work of our method in identifying transcriptional factor binding through analyzing a DNase-seq data set \citep{encode:2012}. We have incorporated the proposed method into the {\tt R} package {\tt MRS}. (This new version of the package is currently available to the public on GitHub and will be uploaed to CRAN after more extensive testing.) 

To close the introduction, we note that while our method is focused on the specific inference problem of multi-group testing, one could also place it in the context of nonparametric hierarchical models for related probability distributions. This is one of the most active areas in Bayesian nonparametric inference. A far-from-exhaustive list of notable examples include the dependent Dirichlet process (DP) and friends \citep{maceachern:1999,maceachern:2000,deiorio:2004,mueller:2004,teh:2006,griffin:2007,rodriguez:2008,dunson:2008,rodriguez:2011,karabatsos:2012}, dependent normalized random measures \citep{rao:2009,griffin:2013,lijoi:2014}, and dependent tail-free processes \citep{jara:2011}. In particular, \cite{deiorio:2004} is among the first to consider related nonparametrically modeling distributions under an ANOVA design. 
By adopting an ANOVA linear model on the location parameters of all mixture kernels in a DP mixture, their model, called ANOVA-DDP, assumes a latent clustering structure in the data and both the cross-group and in-group variations involve shifts in the locations of all clusters. Though ANOVA-DDP is designed for prediction rather than testing and does not incorporate arbitrary forms of variation, we include it in our numerical studies as a comparison because it is the best known and has available software.

\vspace{-1.7em}

\section{Methods}
\label{sec:method}
\vspace{-1.3em}

\subsection{Background}
\label{sec:multiscale_decomp}
\vspace{-0.5em}

We start by providing a brief review of two building blocks for our multi-scale modeling approach to ANDOVA: (i) the Beta-Binomial model \citep{crowder:1978}, and (ii) a multi-scale decomposition of probability distributions based on nested dyadic partitioning of the sample space \citep{ferguson_1973,lavine_1992}. 

{\em The Beta-Binomial model.} Suppose there are $r$ binomial random variables $X_1,X_2,\ldots,X_r$ generated from the following hierarchical model
\vspace{-1.2em}
\begin{align*}
X_j\,|\,p_j \ind {\rm Binomial}(n_j,p_j) \quad \text{and} \quad p_j\,|\,a,b\ind {\rm Beta}(a,b)
\end{align*}
\vspace{-3.2em}

\noindent for $j=1,2,\ldots,r$, which is called the Beta-Binomial (BB) model. A common reparametrization of the model is to let $\theta=a/(a+b)$ and $\nu=a+b$, and so $p_j\,|\,\theta,\nu \ind {\rm Beta}(\theta\nu,(1-\theta)\nu)$. Under this parametrization, $\theta$ is the mean of $p_j$'s and $\nu$ a {\em precision} parameter that controls their variability. The likelihood under the BB model is $L(\theta,\nu) = \prod_{j=1}^{r}D(X_j,n_j-X_j,\theta,\nu)$,
where the mapping $D(\cdot,\cdot,\cdot,\cdot)$ is given by 
\vspace{-1em}
\[
D(n_1,n_2,\theta,\nu)=\begin{cases}B(\theta\nu+n_1,(1-\theta)\nu+n_2)/B(\theta\nu,(1-\theta)\nu) & \text{if $\nu<\infty$}\\
\theta^{n_1}(1-\theta)^{n_2} & \text{if $\nu=\infty$}\end{cases}
\]
\vspace{-2em}

\noindent for $n_1,n_2\in\{0,1,2,\ldots\}$, $\theta\in[0,1]$ and $\nu\in (0,\infty]$ with $B(\cdot,\cdot)$ being the Beta function.

The BB model can be extended to model $k$ groups of Binomial samples. Specifically, let the $i$th group be a collection of $r_i$ Binomial observations $\bX_i = (X_{i1},X_{i2},\ldots,X_{ir_i})$ such that
\vspace{-3.4em}

\begin{align}
\label{eq:bb_kgroup}
X_{ij}\,|\,p_{ij} \ind {\rm Binomial}(n_{ij},p_{ij}), \quad p_{ij}\,|\,\theta_i,\nu\ind {\rm Beta}(\theta_i\nu,(1-\theta_i)\nu). 
\end{align}
\vspace{-3.2em}

\noindent The variations among the $p_{i}$'s are the cross-group variation while that among $p_{ij}$'s for each $i$ the in-group variation. Next we present a multi-scale decomposition of distribution that will allow us to use a collection of BB model to characterize cross-group and in-group variations for data from general probability distributions.

{\em A multi-scale transform of probability distributions.}  Let $Q$ be a Borel probability measure on a sample space $\om$, and let $\T$ be a {\em nested dyadic partition} (NDP) on $\om$. In other words, $\T$ is a collection of sets defined as follows: (i) $\om\in\T$; (ii) for each non-atomic $A\in\T$, (i.e., $A$ contains at least two distinct values,) there exist two nonempty ``children'' sets $A_l$ and $A_r$ in $\T$ that form a partition of $A$, i.e., $A=A_l\cup A_r$ and $A_l\cap A_r=\emptyset$.

When $\om$ is an interval on $\real$, a natural strategy for generating an NDP is to divide each resulting interval $A=(a,b)$ into two subintervals $A_l= (a,c]$ and $A_r=(c,b)$ with $a<c<b$. A simple choice, which we adopt throughout this work for ease of illustration, 
is the middle point $c=(a+b)/2$. A slight generalization is to choose $c$ to be the quantile middle-point between $a$ and $b$, i.e., $c=F^{-1}((F(a)+F(b))/2)$ where $F$ is the cumulative distribution function of some ``base measure'' supported on $\om$, which applies even when $\om$ is unbounded. 

We shall refer to the sets in $\T$ as {\em windows}, and they can be organized on a bifurcating tree. We use $\T^{l}$ to denote the collection of the windows in the $l$th level of the tree---those obtained after $l$ steps of nested partitioning on $\om$. Thus $\T=\cup_{l=1}^{\infty} \T^{l}$. Let $\T^{(K)}=\cup_{l=1}^{K} \T^{l}$ be the collection of all windows up to level $K$.

Given an NDP $\T$, we define a set of coefficients $\{\theta(A):A\in\T\}$ where $\theta(A)=Q(A_l)/Q(A)$ if $Q(A)>0$ and $A$ is non-atomic, and $\theta(A)=1/2$ otherwise. By Carath\'eodory's extension theorem, given that $\T$ generates the Borel $\sigma$-algebra, $Q$ is uniquely determined by these coefficients, and vice versa. 
Thus $\{\theta(A):A\in\T\}$ provides an equivalent representation of $Q$.
Inference tasks regarding $Q$ can thus be formulated in terms of inference on the $\theta(A)$'s. We shall refer to the $\theta(A)$'s as the {\em probability assignment coefficients} (PACs) for $Q$. Each PAC specifies the structure of the distribution at a given scale and location. 

\begin{figure}[t]
  \begin{center}
       \includegraphics[width=0.85\textwidth, clip=TRUE, trim=30mm 20mm 10mm 45mm]{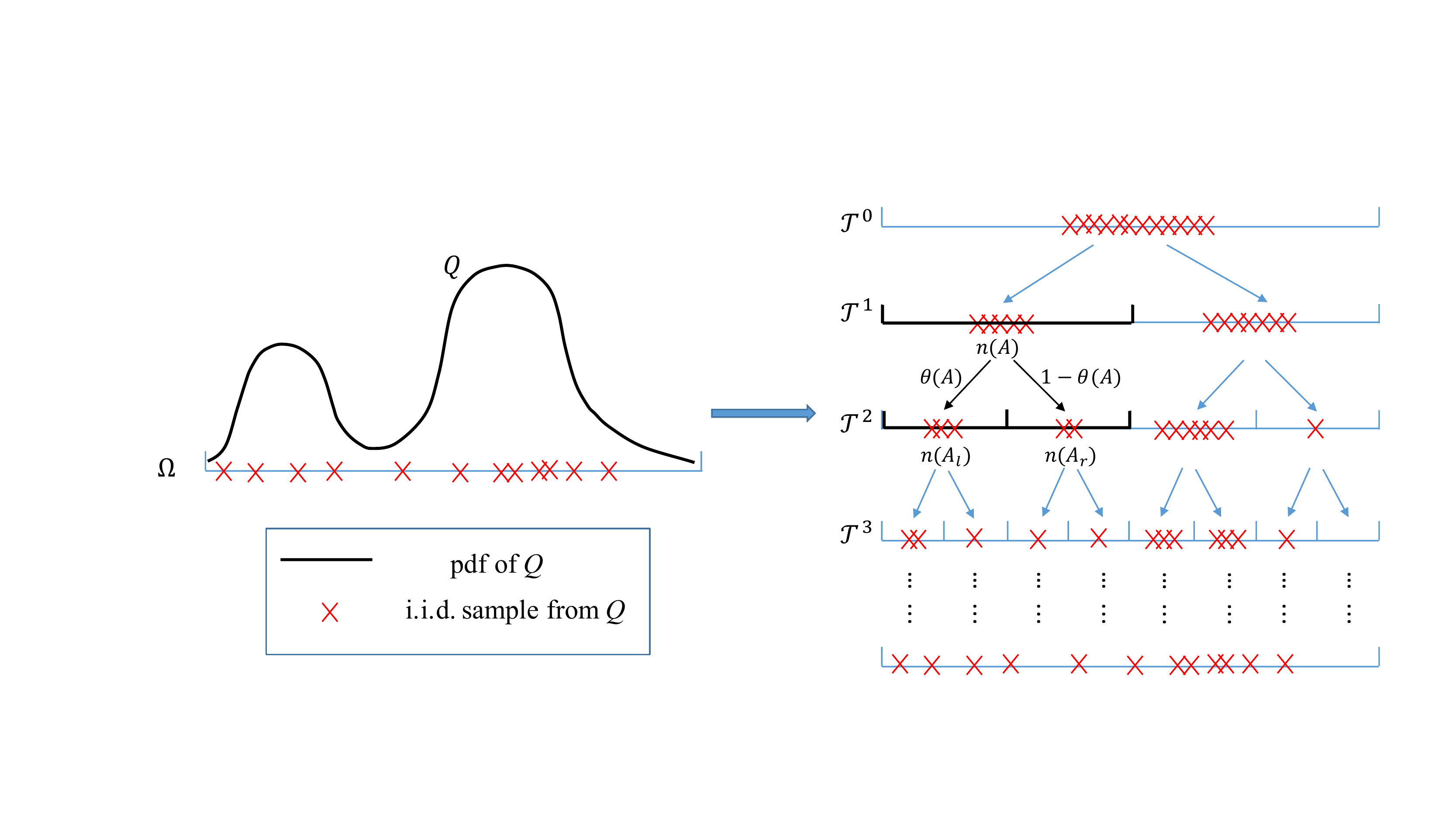}
   \vspace{-1.5em}
    \caption{Multi-scale transform of distributions and window-specific Binomial experiments.}
    \label{fig:mr_decomp}
    \vspace{-1em}
  \end{center}
\end{figure}

In particular, under this decomposition the statistical experiment of observing i.i.d.\ observations from $Q$ is now equivalent to a collection of Binomial experiments, one on each window $A$, given the total number of observations in $A$,
\vspace{-1.5em}

\[
n(A_l)\,|\,n(A),\theta(A)\sim {\rm Binomial}(n(A),\theta(A))
\]
\vspace{-2.5em}

\noindent where $n(A)$ denotes the number of observations in $A$ for all $A\in\T$. Note that $(n(A_l),n(A_r))$ is the sufficient statistic for this window-specific Binomial experiment on $A$. \ref{fig:mr_decomp} illustrates the multi-scale decomposition and the window-specific experiments.

\vspace{-1.2em}

\subsection{A multi-scale Beta-Binomial model for the $k$-group design}
\vspace{-0.7em}

Suppose we have $k$ groups of data, each group consisting of multiple replicate samples collected under a particular condition of interest. For $i=1,2,\ldots,k$, let $r_i$ be the number of data samples (i.e.\ replicates) in Group~$i$. Also, for $j=1,2,\ldots,r_i$, let $\bx_{ij}=(x_{ij1},x_{ij2},\ldots,x_{ij n_{ij}})$ be the $j$th sample collected in Group~$i$, where $n_{ij}$ is the sample size. Suppose the observations $x_{ijl}$'s for $l=1,2,\ldots,n_{ijl}$ are i.i.d.\ from a sampling distribution $Q_{ij}$:
\vspace{-3em}

\begin{align}
\label{eq:sampling_model}
x_{ij1},x_{ij2},\ldots,x_{ij n_{ij}}\,|\,Q_{ij} \iid Q_{ij}.
\end{align}
\vspace{-3em}

Assuming exchangeability among the distributions $Q_{ij}$ within each group,
we model them jointly using a hierarchical model.
Specifically, for $i=1,2,\ldots,k$ and $j=1,2,\ldots,r_i$, let
\vspace{-1em}
\begin{align}
\label{eq:model_ingroup}
\begin{split}
Q_{ij}\,|\,Q_i,\bnu &\iid \pi(Q_{ij}|Q_{i},\bnu) \\
(Q_{1},Q_{2},\ldots,Q_{k}) &\sim \pi(Q_{1},Q_{2},\ldots,Q_{k}),
\end{split}
\end{align}
\vspace{-2em}

\noindent where $Q_i$ is a probability distribution representing the ``centroid'' of the $i$th group, which characterizes the overall common structure shared among the distributions in group $i$. Note that depending on the choice of the model for $Q_{ij}$ given $Q_{i}$, the group ``centroid'' does not have to equal 
the group mean, i.e., $Q_i={\rm E}(Q_{ij}\,|\,Q_i,\bnu)$, although in our following construction, they do coincide. 
Also, $\bnu$ denotes the precision parameters that controls how tightly around the centroids the individual distributions are, i.e., the extent of in-group variation. 

The multi-scale transform of probability distributions motivates a specification of the model in Eq.~\eqref{eq:model_ingroup} in terms of the PACs. Specifically, let $\{\theta_i(A):A\in\T\}$ be the PACs for $Q_i$ and $\{\theta_{ij}(A):A\in\T\}$ those for $Q_{ij}$ for all $i$ and $j$. Then for all $A\in\T$, we let 
\vspace{-1em}
\begin{align}
\label{eq:local_bb}
\theta_{i1}(A),\theta_{i2},\ldots,\theta_{ir_i}(A)\,|\,\theta_i(A),\nu(A) \iid {\rm Beta}(\theta_i(A)\nu(A),(1-\theta_i(A)\nu(A)).
\end{align}
\vspace{-3em}

\noindent A precision parameter $\nu(A)$ is specified for each window $A$, and so $\bnu=\{\nu(A):A\in\T\}$.
Window-specific dispersion parameters allow spatially heterogeneous in-group variation. 

Given the $\theta_{ij}(A)$'s, the statistical experiment of observing each sample $\bx_{ij}$ in Eq.~\eqref{eq:sampling_model} is transformed into a collection of Binomial experiments on $A\in\T$:
\vspace{-1.2em}
\begin{align}
\label{eq:local_binomial}
n_{ij}(A_l)\,|\,n_{ij}(A),\theta_{ij}(A) \sim {\rm Binomial}(n_{ij}(A),\theta_{ij}(A))
\end{align}
\vspace{-3.2em}

\noindent where $n_{ij}(A)$ is the number of observations in $A$ in $\bx_{ij}$.
Therefore we have exactly the BB model in Eq.~\eqref{eq:bb_kgroup} for each window $A$.

In practice, it is infeasible to specify {\em a priori} the proper level of extraneous variation on each $A$. Instead, it is desirable to determine $\nu(A)$ adaptively. The Bayesian strategy to achieving this is by treating each $\nu(A)$ as an unknown parameter with a hyperprior: $\nu(A) \sim G_{A}$.
A simple and suitable choice of the hyperprior is uniform on the $\log_{10}$-scale, i.e., $\log_{10}\nu(A)\sim {\rm Unif}(l,u)$
where lower- and upper-bounds $l$ and $u$ can be chosen to be wide enough to cover common ranges of variation, but not so wide that excessive prior mass is placed in the extremely large or small values. A moderate choice such as $l=-1$ and $u=4$ suffice for most applications. We discuss prior specification in more details in Section~\ref{sec:prior_spec_multiplicity}.

\vspace{-1em}

\subsection{ANDOVA through ms-BB}
\label{sec:ms_ANDOVA}
\vspace{-0.5em}

The goal in ANDOVA is identifying the variation among the group centroids, $Q_1,Q_2,\ldots,Q_k$, while incorporating the extraneous variations. These $k$ distributions are equal if and only if their PACs are equal on each $A$, i.e., they split probability equality between the children of $A$, and their differences can be summarized through finding the windows on which the PACs differ and quantifying how much they differ there. This motivates a window-by-window testing strategy for identifying cross-group variation: 
on each window $A$ (in practice up to some maximum resolution), carry out a test on
\vspace{-1em}
\[
H_0(A): \theta_1(A)=\theta_2(A)=\cdots=\theta_k(A) \quad \text{vs.} \quad H_1(A): \text{otherwise,}
\]
\vspace{-3em}

\noindent and then summarize the evidence from the windows and report the ``significant'' ones, if any.

Classical tests such as (generalized) likelihood ratio test can be directly applied on each~$A$. However we shall consider a Bayesian testing paradigm. This choice is out of practical, rather than philosophical, consideration as it will blend well with the hierarchical modeling framework and later provide us with a convenient means to introducing dependency into the hypotheses for borrowing strength across windows.

Bayesian hypothesis testing on $H_0(A)$ vs $H_1(A)$ requires specifying hyperpriors on the $\theta_i(A)$'s under the two competing hypotheses, and compare the resulting marginal likelihoods (or equivalently the Bayes factor).  To maintain generality, we use $F_{0,A}$ and $F_{1,A}$ to denote the priors for $\theta_i(A)$'s under $H_0(A)$ and $H_1(A)$ respectively. Specifically, under $H_0(A)$ we let
\vspace{-1em}
\begin{align}
\label{eq:f0}
\theta_1(A)=\theta_2(A)=\cdots =\theta_k(A)\sim F_{0,A}
\end{align}
\vspace{-3em}

\noindent and under $H_1(A)$,
\vspace{-3.6em}
\begin{align}
\label{eq:f1}
\theta_i(A)\iid F_{1,A} \quad \text{for $i=1,2,\ldots,k$}.
\end{align}
\vspace{-3em}

\noindent A convenient default choice for the priors $F_{0,A}$ and $F_{1,A}$ is Jeffrey's prior Beta$(0.5,0.5)$, which we shall adopt throughout the rest of the work.

The marginal likelihoods under $H_0(A)$ and $H_1(A)$ are respectively
\vspace{-1em}
\begin{align}
\label{eq:m0}
M_0(A) &=\int\left\{\int \left[\prod_{i=1}^{k}\prod_{j=1}^{r_i}D\left(n_{ij}(A_l),n_{ij}(A_r),\theta,\nu\right)\right]\cdot dF_{0,A}(\theta)\right\} dG_{A}(\nu)
\end{align}
\vspace{-3em}

\noindent and 
\vspace{-2em}
\begin{align}
\label{eq:m1}
M_1(A) &=\int\left\{ \prod_{i=1}^{k}  \int \left[\prod_{j=1}^{r_i}D\left(n_{ij}(A_l),n_{ij}(A_r),\theta_i,\nu\right)\right]\cdot dF_{1,A}(\theta_i)\right\} dG_{A}(\nu)
\end{align}
\vspace{-2em}

\noindent where the mapping $D(\cdot,\cdot,\cdot,\cdot)$ is as defined before.
Bayesian testing can incorporate prior odds for the null vs the alternative through the specification of prior probabilities of the two hypotheses.
To this end, we let $P(H_1(A))=1-P(H_0(A))=\rho(A)\in[0,1]$. 

Bayesian testing can alternatively be expressed in terms of a two-component (``null-and-alternative'') mixture model. Specifically, we introduce an indicator variable $S(A)$ 
such that $S(A)=1$ if $H_0(A)$ is false, and 0 otherwise. 
Then testing $H_0(A)$ vs $H_1(A)$ is essentially inference on the latent indicators $S(A)$ based on the following hierarchical model:
\vspace{-1.5em}
\begin{align*}
S(A)\,|\,\rho(A)&\sim {\rm Bern}(\rho(A))\\
\nu(A)\,|\,G_{A} &\sim G_{A}\\
\theta_1(A),\theta_2(A),\ldots,\theta_k(A)\,|\,S(A) & \sim F_{0,A} \cdot \I_{S(A)=0} + F_{1,A}\times F_{1,A}\times\cdots \times F_{1,A} \cdot \I_{S(A)=1}\\
\theta_{ij}(A)\,|\,\theta_i(A),\nu(A) &\iid {\rm Beta}(\theta_i(A)\nu(A),(1-\theta_i(A))\nu(A)) \\
n_{ij}(A_l)\,|\,n_{ij}(A),\theta_{ij}(A) &\sim {\rm Bin}(n_{ij}(A),\theta_{ij}(A))
\end{align*}
\vspace{-3.5em}

\noindent for all $i=1,2,\ldots,k$ and $j=1,2,\ldots,r_i$. 
We refer to the above hierarchical model as the {\em multi-scale Beta-Binomial} (ms-BB) model. 


Given the data $\bx$, the marginal posterior of $\bS$ is given by 
\vspace{-1.5em}
\begin{align*}
S(A)\,|\,\brho,\balp,\bx&\sim {\rm Bern}(\trho(A)),
\end{align*}
\vspace{-3.5em}

\noindent where $\trho(A)$ is the posterior probability for $H_1(A)$, and adopting convention in Bayesian model choice, we shall call it the {\em posterior marginal alternative probability} (PMAP) on window $A$:
\vspace{-0.5em}
\begin{align}
\label{eq:pmap}
\trho(A) = {\rm PMAP}(A)=\frac{\rho(A)M_1(A)}{(1-\rho(A))M_0(A)+\rho(A)M_1(A)}=\frac{\rho(A)\cdot{\rm BF}(A)}{(1-\rho(A))+\rho(A)\cdot {\rm BF}(A)}
\end{align}
\vspace{-2.5em}

\noindent where ${\rm BF}(A)=M_1(A)/M_0(A)$ is the Bayes factor (BF) for $H_0(A)$ vs $H_1(A)$. The PMAP summarizes the evidence, both prior and empirical, for the presence of cross-group variation in $\theta_i(A)$. We can call a window ``significant'', that is, on which $H_0(A)$ is to be rejected, when the PMAP is large. (See Section~\ref{sec:prior_spec_multiplicity} for details on how to choose a threshold.) 

It is often also desirable to quantify the extent of cross-group variation on those windows identified as significant. To this end, we introduce a notion of ``effect size'' that measures the extent to which each group centroid $Q_i$ differs from the rest of the centroids on each~$A$. While many options are available, we consider the log odds ratio between $\theta_i(A)$ from that of $\bar{\theta}_{(-i)}(A):=\sum_{i'\neq i} \theta_{i'}(A)/(k-1)$. That is,
\vspace{-1em}
\[
{\rm eff}_{i}(A) = \log \frac{\theta_i(A)/(1-\theta_i(A))}{\bar{\theta}_{(-i)}(A)/(1-\bar{\theta}_{(-i)}(A))}.
\]
\vspace{-2.5em}

\noindent We can summarize/estimate the posterior effect size using, say, the posterior expectation
\vspace{-1em}
\begin{align}
\label{eq:post_effect_size}
{\rm E}({\rm eff}_{i}(A)\,|\,\bx)={\rm PMAP}(A) \times {\rm E}({\rm eff}_{i}(A)\,|\,S(A)=1,\bx).
\end{align}
\vspace{-3em}

\noindent We provide a recipe for evaluating ${\rm E}({\rm eff}_{i}(A)\,|\,S(A)=1,\bx)$ later in Eq.~\eqref{eq:post_effect_size2}.
\vspace{-1em}

\subsection{Computational strategies}
\vspace{-0.5em}

ANDOVA under the ms-BB model requires the evaluation of a number of integrals. In particular, for calculating the PMAPs, the marginal likelihood terms $M_0(A)$ and $M_1(A)$ are needed and for the effect size, its posterior expectation. One can appeal to either posterior sampling or numerical approximation. Here we shall opt for the latter, because (i) the BB (log-)likelihood is unimodal and strictly concave and thus numerical integration achieves high accuracy and computational efficiency
and (ii) the numerical approach works well with the message passing strategy we present later, when incorporating dependency among windows. 
 
Two of the most simple strategies for evaluating the integrals are Laplace (mode-based) approximation and numerical quadrature, in particular a finite Riemann (grid-based) approximation. 
The term $D(n_{ij}(A_l),D_{ij}(A_r),\theta,\nu)$ in the integrand in Eqs.~\eqref{eq:m0} and \eqref{eq:m1} is the likelihood of a BB model, whose log is unimodal and strictly concave in $\theta$, and reasonably well approximated by a Gaussian density. 
Thus Laplace approximation is a suitable choice for integration over $\theta_i(A)$ given $\nu(A)$. On the other hand, the marginal likelihood in the precision parameter $\nu(A)$ is typically quite flat and so finite Riemann integration can be very effective in achieving high numerical precision without requiring a dense grid. This suggests that
a hybrid strategy combining both integration methods can be highly effective for evaluating $M_0(A)$ and $M_1(A)$. 
Specifically, we evaluate the inner integration on $\theta_i(A)$'s given $\nu(A)$ using Laplace approximation and the outer integration on the precision parameter $\nu(A)$ using finite Riemann integral.  We next provide details on each. 

{\em Laplace approximation on the inner integral.} The inner integral for $M_0(A)$ is
\vspace{-1em}
\begin{align*}
L_{0,A}(\nu)=&\int \left[\prod_{i=1}^{k}\prod_{j=1}^{r_i}D\left(n_{ij}(A_l),n_{ij}(A_r),\theta,\nu\right)\right]\cdot dF_{0,A}(\theta)\\
= &\int\exp\left\{\sum_{i,j} \log D\left(n_{ij}(A_l),n_{ij}(A_r),\theta,\nu\right)+\log f_{0,A}(\theta)\right\} d\theta\\	
= &\int \exp\{h_{0,\nu}(\theta)\}d\theta\\
\approx&\exp\{h_{0,\nu}(\hat{\theta}_{0,\nu})\}\cdot\sqrt{-2\pi/h_{0,\nu}''(\hat{\theta}_{0,\nu})}=\hat{L}_{0,A}(\nu)
\end{align*}
\vspace{-3em}

\noindent where $f_{0,A}=dF_{0,A}/d\theta$, $h_0(\theta,\nu)=\sum_{i,j} \log D\left(n_{ij}(A_l),n_{ij}(A_r),\theta,\nu\right)+\log f_{0,A}(\theta)$, and $\hat{\theta}_{0,\nu}={\rm argmax}_{\theta}\, h_{0,\nu}(\theta)$. The optimization can be carried out using Newton-Raphson, as the first and second derivatives of $\log D\left(n_{ij}(A_l),n_{ij}(A_r),\theta,\nu\right)$ can be expressed in terms of digamma and trigamma functions, and $\log f_{0,A}(\theta)=-\frac{1}{2}\log\theta -\frac{1}{2}\log(1-\theta)$ for $F_{0,A}$ being Beta(0.5,0.5). 

Computing $M_1(A)$ requires the evaluation of $k$ one-dimensional inner integrals, one for each of the $k$ groups. 
For the $i$th group, the corresponding inner integral is 
\vspace{-0.7em}
\begin{align*}
L_{i,A}(\nu)&=\int \left[\prod_{j=1}^{r_i}D\left(n_{ij}(A_l),n_{ij}(A_r),\theta,\nu\right)\right]\cdot dF_{1,A}(\theta)\\
&\approx \hat{L}_{i,A}=\exp\{h_{i,\nu}(\hat{\theta}_{i,\nu})\}\cdot \sqrt{-2\pi/h_{i,\nu}''(\hat{\theta}_{i,\nu})}
\end{align*}
\vspace{-3.2em}

\noindent where $h_{i,\nu}(\theta)=\sum_{j=1}^{r_i}\log D(n_{ij}(A_l),D_{ij}(A_r),\theta,\nu)+\log f_{1,A}(\theta)$ and $\hat{\theta}_{i,\nu}={\rm argmax}_{\theta}h_{i,\nu}(\theta)$. 
\vspace{0.5em}

{\em Finite Riemann approximation for the outer integral.} Given the inner integral(s) evaluated at a grid of $\nu$ values, $\nu^{(0)},\nu^{(1)},\ldots,\nu^{(T)}$, we have
\vspace{-1em}
\begin{align*}
M_0(A) &\approx \sum_{h=1}^{T} \hat{L}_{0,A}(\nu^{(h)})(G_{A}\left(\nu^{(h)})-G_{A}(\nu^{(h-1)})\right)\\
\text{and} \qquad M_1(A) &\approx\sum_{h=1}^{T} \left[\prod_{i=1}^{k}\hat{L}_{i,A}(\nu^{(h)})\right]\left(G_{A}(\nu^{(h)})-G_{A}(\nu^{(h-1)})\right).
\end{align*}
\vspace{-2.5em}

\noindent These marginal likelihoods allow us to evaluate PMAP$(A)$ according to Eq.~\eqref{eq:pmap}. 

Following the same strategy, ${\rm E}({\rm eff}_{i}(A)|S(A)=1,\bx)$ can be evaluated as
\vspace{-0,5em}
\begin{align}
\label{eq:post_effect_size2}
\begin{split}
&{\rm E}({\rm eff}_{i}(A)|S(A)=1,\bx)=\int {\rm eff}_{i}(A) \cdot \left[ \prod_{i'=1}^{k} L_{i',A}(\nu) \right]\cdot\frac{1}{M_1(A)} \,dG_{A}(\nu)\\
&\!\!\approx\frac{1}{M_1(A)}\sum_{h=1}^{T}\left(\log \frac{\hat{\theta}_{i,\nu^{(h)}}(A)/(1-\hat{\theta}_{i,\nu^{(h)}}(A))}{\bar{\hat{\theta}}_{(-i),\nu^{(h)}}(A)/(1-\bar{\hat{\theta}}_{(-i),\nu^{(h)}}(A))}\right) \left[\prod_{i'=1}^{k}\hat{L}_{i',A}(\nu^{(h)})\right](G_{A}(\nu^{(h)}) - G_{A}(\nu^{(h-1)})).
\end{split}
\end{align}
\vspace{-2.5em}

\noindent with $\bar{\hat{\theta}}_{(-i),\nu}=\sum_{i'\neq i} \hat{\theta}_{i',\nu}(A)/(k-1)$. Combining this with the PMAP, we can compute the posterior expected effect size on each $A$, ${\rm E}({\rm eff}_i(A)\,|\,\bx)$, following Eq.~\eqref{eq:post_effect_size}.

\vspace{0.5em}

{\em Full posterior of the ms-BB model.} It is worth noting, though not needed in this work, that in addition to ANDOVA, one could also use the ms-BB as a general purpose hierarchical Bayesian model for $k$-groups of related probability measures. 
General Bayesian inference can proceed based on sampling from the full posterior of the model. To describe the full posterior, we give the conditional posterior of all the other unknown parameters given the latent indicators $S(A)$:
\vspace{-2em}
\begin{align*}
\nu(A)\,|\,S(A),G_{A},\brho,\balp, \bx &\sim \tilde{G}_{0,A}\cdot \I_{S(A)=0} + \tilde{G}_{1,A}\cdot \I_{S(A)=1}\\
\theta_1(A)=\theta_2(A)=\cdots = \theta_{k}(A) \,|\,S(A)=0,\nu(A)=\nu,\bx &\sim \tilde{F}_{0,A|\nu}\\
\theta_i(A)\,|\,S(A)=1,\nu(A)=\nu,\bx &\ind \tilde{F}^{(i)}_{1,A|\nu}\\
\theta_{ij}(A)\,|\,\theta_i(A),\nu(A),\bx &\ind {\rm Beta}(\tilde{\theta}_{ij}(A)\tilde{\nu}_{ij}(A),(1-\tilde{\theta}_{ij}(A))\tilde{\nu}_{ij}(A)),
\end{align*}
\vspace{-3em}

\noindent where $\tilde{\theta}_{ij}(A) = (\theta_i(A)\nu(A)+n_{ij}(A_l))/(\nu(A)+n_{ij}(A))$, $\tilde{\nu}_{ij}(A) = \nu(A)+n_{ij}(A)$, and
\vspace{-1em}
\begin{align*}
d\tilde{G}_{0,A}(\nu) &= L_{0,A}(\nu) \cdot dG_{A}(\nu)/M_0(A)\approx \hat{L}_{0,A}(\nu) \cdot dG_{A}(\nu)/M_0(A)\\
d\tilde{G}_{1,A}(\nu) &= \left(\prod_{i=1}^{k}  L_{i,A}(\nu)\right) \cdot dG_{A}(\nu)/M_1(A)\approx \left(\prod_{i}^{k}\hat{L}_{i,A}(\nu)\right) \cdot dG_{A}(\nu)/M_1(A)\\
d\tilde{F}_{0,A|\nu}(\theta) &\propto \prod_{i=1}^{k}\prod_{j=1}^{r_i}D\left(n_{ij}(A_l),n_{ij}(A_r),\theta,\nu\right)\cdot dF_{0,A}(\theta)\approx {\rm N}\left(\hat{\theta}_{0,\nu},1/h''_{0,\nu}(\hat{\theta}_{0,\nu})\right)\\
d\tilde{F}^{(i)}_{1,A|\nu}(\theta) &\propto \prod_{j=1}^{r_i} D\left(n_{ij}(A_l),n_{ij}(A_r),\theta,\nu\right)\cdot dF_{1,A}(\theta)\approx {\rm N}\left(\hat{\theta}_{i,\nu},1/h''_{i,\nu}(\hat{\theta}_{i,\nu})\right).
\end{align*}
\vspace{-3em}

Sampling from $\tilde{G}_{0,A}$ and $\tilde{G}_{1,A}$ can be based on the discrete approximation on the grid points $(\nu^{(1)},\nu^{(2)},\ldots,\nu^{(T)})$ used in evaluating the Riemann integrals. Sampling from $\tilde{F}_{0,A|\nu}$ and $\tilde{F}_{1,A|\nu}$ can be achieved based on the above normal approximations. All the necessary quantities for posterior sampling are available as part of evaluating the marginal likelihoods.

\vspace{-1.2em}

\subsection{Graphical ms-BB for enhanced ANDOVA} 
\label{sec:graphical_ms_ANDOVA}
\vspace{-0.7em}

The above ms-BB model treats the inference problem on each window $A$ independently (given the hyper-parameters). 
Thus the resulting test of $H_0(A)$ is ``window-autonomous'', in the sense that no information (beyond that regards the hyperparameters) is shared among the windows. Such a strategy can perform well for windows with high ``signal-to-noise ratio'', i.e., either the effect size is large or the window contains large amounts of data.

In many modern applications, however, one does not have either, especially when the underlying cross-group variations is of a local nature---involving only a very small portion of the data/probability mass. Small windows that target such local variations contain limited data, with some replicate samples potentially not even present in such windows. Consequently, the window-autonomous inferential strategy is typically underpowered on those windows. 

One remedy is to borrow strength among the windows. The motivation is that distributional variations tend to occur in a dependent manner both across space and across scales. Nearby and nested windows in the location-scale tree tend to contain cross-group variations simultaneously, and such dependency is particularly strong between a parent window and its children. 
Accordingly, the null hypotheses $H_0(A)$ on nearby/nested scanning windows are often true or false together, and such windows contain empirical evidence for or against $H_0(A)$ in a correlated manner. Incorporating this dependency will allow us to borrow information across the windows, thereby enhancing the discriminatory power on each $H_0(A)$.

Our hierarchical modeling approach to ANDOVA affords a natural means to inducing such dependency across the windows---through an additional layer of modeling, in the form of a graphical model, that links the $S(A)$'s. A particularly simple and yet flexible graphical model that suffices is a Bayesian network called the Markov tree (MT) \citep{crouse_etal_1998}. Not only does it induce the desired spatial-scale dependency, its tree structure also ensures that the posterior marginal probabilities, here the PMAPs, can still be calculated exactly using message passing \citep{willsky:2001}. This graphical modeling strategy for inducing spatial-scale dependency has been shown to be highly effective in the context of two-sample comparison \citep{soriano&ma:2015}. 

Specifically, under an MT model for $\{S(A):A\in\T\}$, the $S(A)$'s are generated through top-down Markov transitions.
In particular, for any window $A$, let $A_p$ be the parent of $A$ in the previous scale. Then $S(A)$ is drawn conditional on the value of $S(A_p)$
\vspace{-1.7em}

\[
P(S(A)=s'\,|\,S(A_p)=s) = \rho_{s,s'}(A)
\]
\vspace{-2.7em}

\noindent where $\rho_{s,s'}(A)$ for $s,s'\in\{0,1\}$ are referred to as the {\em null/alternative transition probabilities}, which can be organized into a $2\times 2$ transition matrix 
\vspace{-1em}
\[ \brho(A)= \left(
\begin{array}{cc}
 \rho_{0,0}(A) & \rho_{0,1}(A) \\
 \rho_{1,0}(A) & \rho_{1,1}(A)     
\end{array}\right). \]
\vspace{-2em}

\noindent In the following, let $\brho=\{\brho(A):A\in\T\}$ and we write $\S\,|\,\brho\sim {\rm MT}(\brho)$. The sample space $\om$ has no parent, and so we draw $S(\om)$ from an unconditional Bernoulli prior just as before. To unify the notation, we also define a transition matrix for $\om$ with two equal rows $\rho_{0,s}(\om)=\rho_{1,s}(\om)=P(S(\om)=s)$ for $s=0,1$. The full hierarchical model now becomes
\vspace{-1.5em}
\begin{align*}
\S\,|\,\brho&\sim {\rm MT}(\brho)\\
\nu(A)\,|\,G_{A} &\sim G_{A}\\
\theta_1(A),\theta_2(A),\ldots,\theta_k(A)\,|\,S(A) & \sim F_{0,A} \cdot \I_{S(A)=0} + F_{1,A}\times F_{1,A}\times\cdots \times F_{1,A} \cdot \I_{S(A)=1}\\
\theta_{ij}(A)\,|\,\theta_i(A),\nu(A) &\iid {\rm Beta}(\theta_i(A)\nu(A),(1-\theta_i(A))\nu(A)) \\
n_{ij}(A_l)\,|\,n_{ij}(A),\theta_{ij}(A) &\sim {\rm Bin}(n_{ij}(A),\theta_{ij}(A))
\end{align*}
\vspace{-3.5em}

\noindent for all $i=1,2,\ldots,k$ and $j=1,2,\ldots,r_i$. 

ANDOVA under this graphical ms-BB model requires the computation of the PMAPs, which is available from recursive message passing. Instead of presenting the computational recipe in an algorithmic fashion, however, we present its output---the marginal posterior of the latent states $\S$ in the following theorem.

\begin{thm}
\label{thm:posterior}
Under the above hierarchical model,
given i.i.d.\ observations from the $Q_{ij}$'s, the marginal posterior of $\S$ is still an MT given by
\vspace{-3.6em}

\begin{align*}
\S\,|\,\brho,\bx &\sim {\rm MT}(\tilde{\brho})
\end{align*}
\vspace{-4.1em}

\noindent with 
$\tilde{\brho}=\{\tbrho(A):A\in\T\}$ given as
\vspace{-1.2em}
\[
\tilde{\brho}(A) = {\rm diag}(\bphi(A))^{-1}\cdot \brho(A)\cdot 
 {\rm diag}(\vect{m}(A))
\cdot {\rm diag}\left(\bphi(A_l) \circ \bphi(A_r)\right)
\]
\vspace{-3.2em}

\noindent where 
$\vect{m}(A)=(1,{\rm BF}(A))'$, $\circ$ is the Hadamard product, and $\bphi:\T \rightarrow \real^{2}$ is as follows
\vspace{-1em}
\begin{align*}
\bphi(A) = \begin{cases}
\brho(A) \cdot  {\rm diag}\left(\bphi(A_l) \circ \bphi(A_r)\right) 
\cdot \vect{m}(A)& \text{if $\sum_{j} n_{ij}(A) > 0$ \text{for at least two $i$'s and $A$ has children}}\\
(1,1)' & \text{otherwise.}
\end{cases}
\end{align*}
\end{thm}

\noindent Remark: This theorem applies to both infinitely deep partition trees and finite ones, i.e., those that have a maximum resolution. In the latter case, the windows in the maximum resolution have no children.
\vspace{0.5em}

Theorem~\ref{thm:posterior} provides an analytical recipe for evaluating the marginal posterior of $\S$ through recursive message passing. The expression for the posterior transition matrix $\tilde{\rho}(A)$ consists of four interpretable pieces: ${\rm diag}(\bphi(A))^{-1}$ is a normalizing matrix to ensure that the resulting matrix is a valid transition matrix (each row sums to 1); $\brho(A)$ incorporates the prior information; $\vect{m}(A)$ contains the empirical evidence from the data on window $A$ in support of the $H_0(A)$ vs $H_1(A)$; and ${\rm diag}(\bphi(A_l)\circ \bphi(A_r))$ contains the information, or the ``message'', from the children of $A$ regarding what finer scale evidence within $A$ reveal about $H_0(A)$ and $H_1(A)$. Intuitively, if all of the observations in $A$ come from samples in a single group, then there is no empirical evidence in $A$ that favors $H_0(A)$ or $H_1(A)$, and so the message to $A$ from its descendants is $\phi(A)=(1,1)'$. One can check that on those nodes $\tilde{\brho}(A)=\brho(A)$, and so the prior and posterior transition matrices coincide. 
Operationally, the theorem suggests that one can initiate a message either from the leaf nodes in $\T$ or from those nodes with one observations exclusively from one group, and recursively pass the evidence back to the root of $\T$. 

Based on the marginal posterior of the $S(A)$'s, we can calculate the PMAP on each window $A$. In addition, if one carries out the testing up to a maximum resolution of $K$, we can also compute the posterior {\em joint} alternative probability (PJAP), or the posterior probability for $S(A)=1$ on at least one window. 
Specifically for each $A\in\T$, let $\vect{\tvarphi}(A)=\left(1-{\rm PMAP}(A),{\rm PMAP}(A)\right)'$.
Then for $A=\om$
\vspace{-1.2em}
\[
\vect{\tvarphi}(\om) = (\trho_{0,0}(\om),\trho_{0,1}(\om))'.
\]
\vspace{-3.2em}

\noindent Now suppose $\{\vect{\tvarphi}(A):A\in \T^{(k)}\}$ have been computed for some $k\geq 0$, then for any $A\in \T^{k+1}$,
\vspace{-1.5em}
\[
\vect{\tvarphi}(A) = \tilde{\brho}(A)' \vect{\tvarphi}(A_p)
\]
\vspace{-2.5em}

\noindent where as before, $A_p$ is the parent of $A$ in $\T^{k}$. In addition, the PJAP for all windows up to resolution $K$ 
is given by
\vspace{-1.2em}
\[
{\rm PJAP} = P(S(A)=1\text{ for some } A\in \T^{(K)}\,|\,\bx)=1-\prod_{A\in\T^{(K)}} \trho_{0,0}(A).
\]
\vspace{-2.5em}

In addition, once given the $S(A)$'s, the conditional posterior of the graphical ms-BB model stays the same as before. The posterior expected effect size on $A$ is still given by Eq.~\eqref{eq:post_effect_size}.

\vspace{-1em}

\subsection{Prior specification and rules for calling significant windows}
\label{sec:prior_spec_multiplicity}
\vspace{-0.5em}

We have introduced a general approach to ANDOVA using the graphical ms-BB model, but have so far left out two important questions that will arise in actual applications. The first is how to specify the hyperparameters, and the other is how to choose the threshold for the PMAPs in calling ``significant'' findings. We consider them together in this subsection as we use the same guiding principles---in terms of multiple testing control---to address each.
\vspace{0.5em}

{\em Prior specification.} First we consider the choice of hyperparameters. In particular, we need to determine the prior transition matrices $\brho(A)$. In most applications, {\em a priori} it is reasonable to treat the windows in each resolution symmetrically by specifying $\brho(A)$ as a function the level of the window $A$. Suppose now we let $\brho(A)=\brho(j)$ for all $A$ in level $j$. 

To specify $\brho(j)$, we note that $\rho_{0,1}(j)$ is the prior probability for a window $A$ to have $H_1(A)$ being true while that for its parent is not. 
Because the number of windows grow at a rate of $2^j$ with the resolution, an appropriate choice is $\rho_{0,1}(j)=\beta 2^{-j}$, which counters the growth in the number of windows and keep constant the prior expected number of ``alternative'' windows. 
One way to elicit $\beta$ is by setting the prior joint alternative probability (PrJAP), 
\vspace{-1em}
\[ 
{\rm PrJAP}=P(S(A)=1\text{ for some $A\in\T^{(K)}$})=1-\prod_{j=1}^{K}(1-\beta 2^{-j})^{2^{j}}
\]
\vspace{-3em}
 
\noindent to some desired level such as 50\%.

On the other hand, $\rho_{1,1}(j)$ is the prior probability for a window to ``inherit'' a signal from its parent, which characterizes the extent of dependency among the windows. Such dependency typically does not decay with the level and so a parsimonious choice is to let $\rho_{1,1}(j)=\delta \in [0,1]$ for all levels. The value of $\delta$ can be elicited using the prior expected proportion of significant windows, that is the proportion of windows on which $H_0(A)$ is false. To this end, note that we can get the prior marginal alternative probability (PrMAP) for each window $A$ using in the same way as we compute the PMAPs---simply replacing $\vect{\tvarphi}(A)$ with $\vect{\varphi}(A)=({\rm PrMAP}(A),1-{\rm PrMAP}(A))'$ and $\tilde{\brho}(A)$ with $\brho(A)$ in the recursive derivation for $\vect{\tvarphi}(A)$ in Section~\ref{sec:graphical_ms_ANDOVA}. The sum of PrMAPs over all the windows up to level $K$ gives the total prior expected number of alternative windows, which given $\beta$ is a monotone function in $\delta$, and so we can choose $\delta$ to set this expected number of ``signals'', or equivalently, the proportion of signals among all windows, at a desired level.

{\em Decision rules for calling significant windows.} After computing the PMAPs,
one can adopt the simple decision rule that call windows ``significant'' if the PMAP is larger than a threshold $c$. 
Such rules are commonly adopted in Bayesian model choice, see for example \cite{barbieri2004} where a threshold of $c=0.5$ is recommended. More generally, the threshold can be chosen to control for multiple testing in terms of the (Bayesian) false discovery rate (FDR) \citep{muller_etal_2007}. To this end, note that if we call the windows with PMAP$(A)>c$ as significant, then the Bayesian FDR, defined as the ratio of the posterior expected number of false positives (NPR) divided by the total number of positives, is
\vspace{-0.8em}
\[
{\rm FDR}(c) = 1- \frac{\sum_{A:{\rm PMAP}(A)>c} {\rm PMAP}(A)}{|\{A:{\rm PMAP}(A)>c\}|},
\]
\vspace{-3em}

\noindent which can be computed for any given threshold $c$. We can also choose the threshold to achieve any desired Bayesian FDR, such as 10\%. 

\vspace{-1em}

\section{Numerical examples}
\label{sec:numerical_examples}
\vspace{-0.8em}

We carry out a simulation study to investigate the performance of our ANDOVA method based on the ms-BB model under a number of representative scenarios, and compare it to and several other methods, namely ANOVA-DDP and a few state-of-the-art $k$-sample comparison methods. For simplicity, from now on, we shall use ``ANDOVA'' to refer specifically to our ms-BB based method.
The results demonstrate several important features of ANDOVA in contrast to the other methods. First, in the presence of confounders, the test statistics for the $k$-sample methods---both the $p$-value under the frequentist paradigm or the posterior probability of the null under Bayesian testing---lose their meaning when the null hypothesis is true (i.e., when there is no cross-group difference). This can result in large numbers of false positives if one interprets the test statistic by face value. In contrast, by taking into account the in-group variation, the meaning of the test statistic is restored under ANDOVA. Second, even if one knows the true behavior of the test statistic under the null, and has adjusted its interpretation accordingly, ANDOVA can still substantially improve the performance by taking into account the in-group variation. Finally, ANDOVA under the graphical ms-BB allows us to identify the  structure of the cross-group difference rather than just testing its existence. We demonstrate each of these in Sections~\ref{sec:null_behavior}, \ref{sec:testing_performance}, and \ref{sec:identifying_difference}. In all of the examples throughout this section, we adopt a maximum of 12 resolution levels (i.e., $K=11$), and let $\beta=0.07$ and $\delta=0.4$ to set the prior joint alternative (or null) probability at 50\% and the prior expected number of signals at all levels combined at about 2 following Section~\ref{sec:prior_spec_multiplicity}.

\vspace{-1em}

\subsection{Null behavior of cross-sample tests under the $k$-group setting}
\label{sec:null_behavior}
\vspace{-0.5em}

We begin by investigating the behavior of popular $k$-sample tests under the null---that is no cross-group variation---in the presence of extraneous, in-group variation. 
Specifically, we simulate two groups of data sets with each group consisting of four replicate samples. Both groups share the same centroid distribution and all differences among the replicate samples are due to in-group variation. (While our framework works generally for $k$-group designs, we set $k=2$ because the available software for popular cross-sample comparison methods that we shall compare to are for two-sample comparisons.) 

In our simulations, the true sampling distribution of each replicate sample is a mixture distribution with three mixing components characterizing features of different scales. The two group ``centroid'' distribution are both
\vspace{-0.8em}
\begin{align}
\label{eq:null_centroid}
\frac{1}{3} \, {\rm N}(1,0.05^2) + \frac{1}{3}\, {\rm N}(1.5,0.2^2) + \frac{1}{3}\, {\rm N}(2.5,0.1^2).
\end{align}
\vspace{-3em}

\noindent For each group, we introduce in-group variation by adding perturbation into the mixture weights. Specifically, the weights for the $j$th sample in the $i$th group are generated from a baseline logit model with random intercepts by drawing three independent standard normal variables $Z_{ij1},Z_{ij2},Z_{ij3}$, with the corresponding weights $(\pi_{ij1},\pi_{ij2},\pi_{ij3})$ given by
\vspace{-0.8em}
\[
\pi_{ijl} = \frac{e^{Z_{ijl}}}{\sum_{l'=1}^{3}e^{Z_{ijl'}}} \quad \text{for $l=1,2,3$.}
\]
\vspace{-2.4em}

\noindent Then the observation in that sample is drawn from the following distribution
\vspace{-1.2em}
\[
Q_{ij}: \pi_{ij1} \, {\rm N}(1,0.05^2) + \pi_{ij2}\, {\rm N}(1.5,0.2^2) + \pi_{ij3}\, {\rm N}(2.5,0.1^2).
\]
\vspace{-3.2em}

\begin{figure}[t]
\begin{center}
\includegraphics[width=28em, clip=TRUE, trim = 0 5mm 0 15mm]{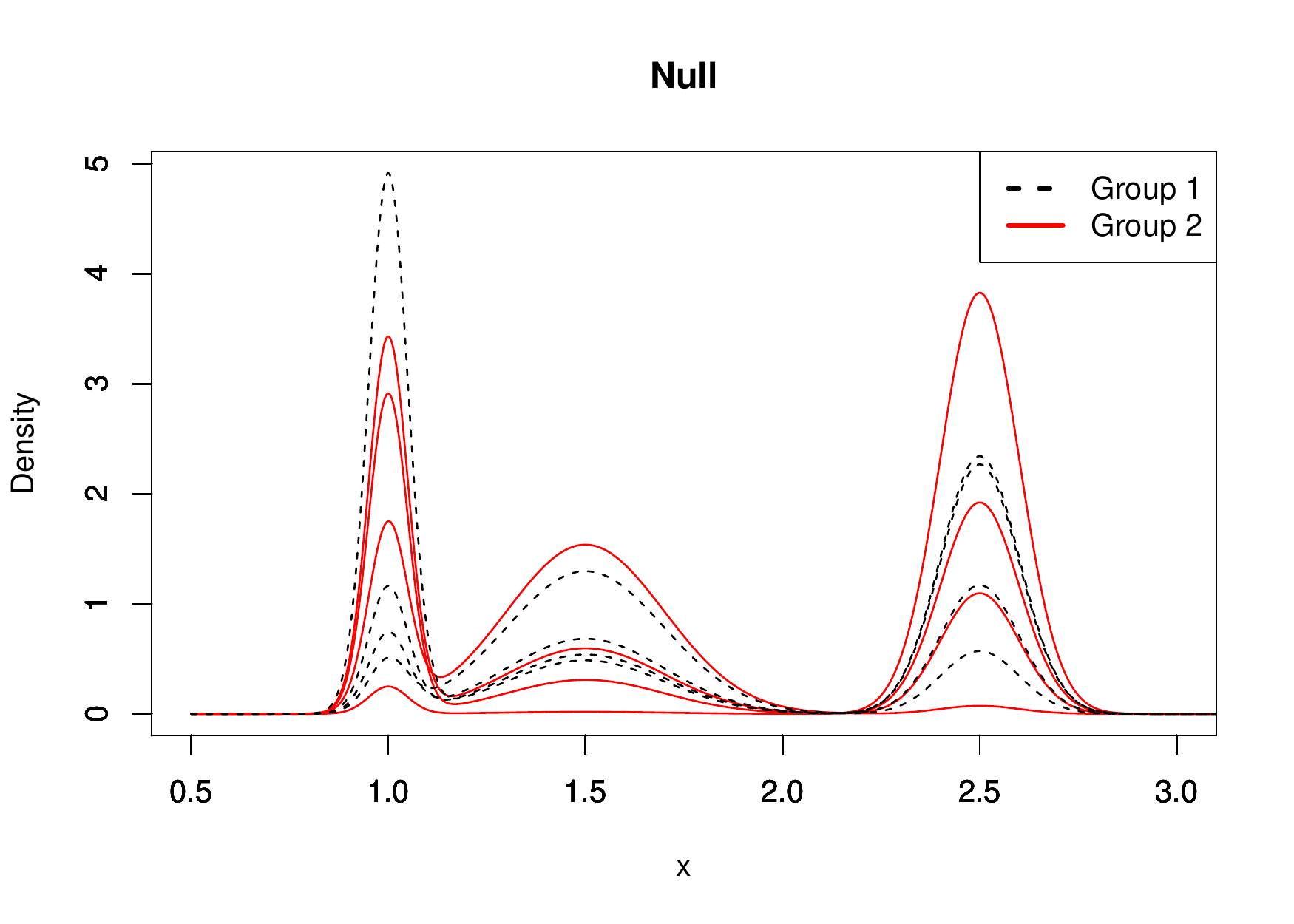}
\vspace{-0.5em}
\caption{A realization of the sampling distributions under the null simulation scenario.}
\label{fig:null_true_den}
\end{center}
\vspace{-0.5em}
\end{figure}

\noindent \ref{fig:null_true_den} shows a realization of the eight sampling distributions (four in each group).

Two things are worth mentioning: First, this underlying true model we simulate from is not the ms-BB model. In particular, the mechanism in which we introduce in-group variation is different. We feel that the comparison is more fair to the competitors if we do not simulate from our own model. Second, our method does not utilize the mixture structure of the distributions. 

A total number of 500 observations (with all four samples combined) are generated for each of the two groups. In each simulation run, the total sample size of 500 for each group is randomly split among the four replicate samples according to a multinomial distribution whose class probability vector is drawn from a Dirichlet$(1,1,1,1)$.

We carry out the simulation 500 times and each time we apply the following methods: (i) ANDOVA using graphical ms-BB, (ii) ANOVA-DDP \citep{deiorio:2004}, (iii) Cramer test \citep{baringhaus_franz_2004}, (iv) the $k$-nearest neighbor (KNN) test \citep{schilling_1986,henze_1988}, and (v) a Bayesian nonparametric two-sample test based on the P\'olya tree (PT) \citep{holmes_etal_2012}, as well as (vi) a generalization of the PT test \citep{chen_hanson_2012}, which we shall refer to as the CH test. In addition, to understand how incorporating in-group variation impacts the performance of our approach, we also include into the comparison (vii) a ``$k$-sample'' version of the graphical ms-BB model restricted to having $\nu(A)= \infty$ for all $A$ and thus no in-group variation---all replicates in each group have exactly the same distribution.

\begin{figure}[htbp]
\begin{center}
\includegraphics[width=40em]{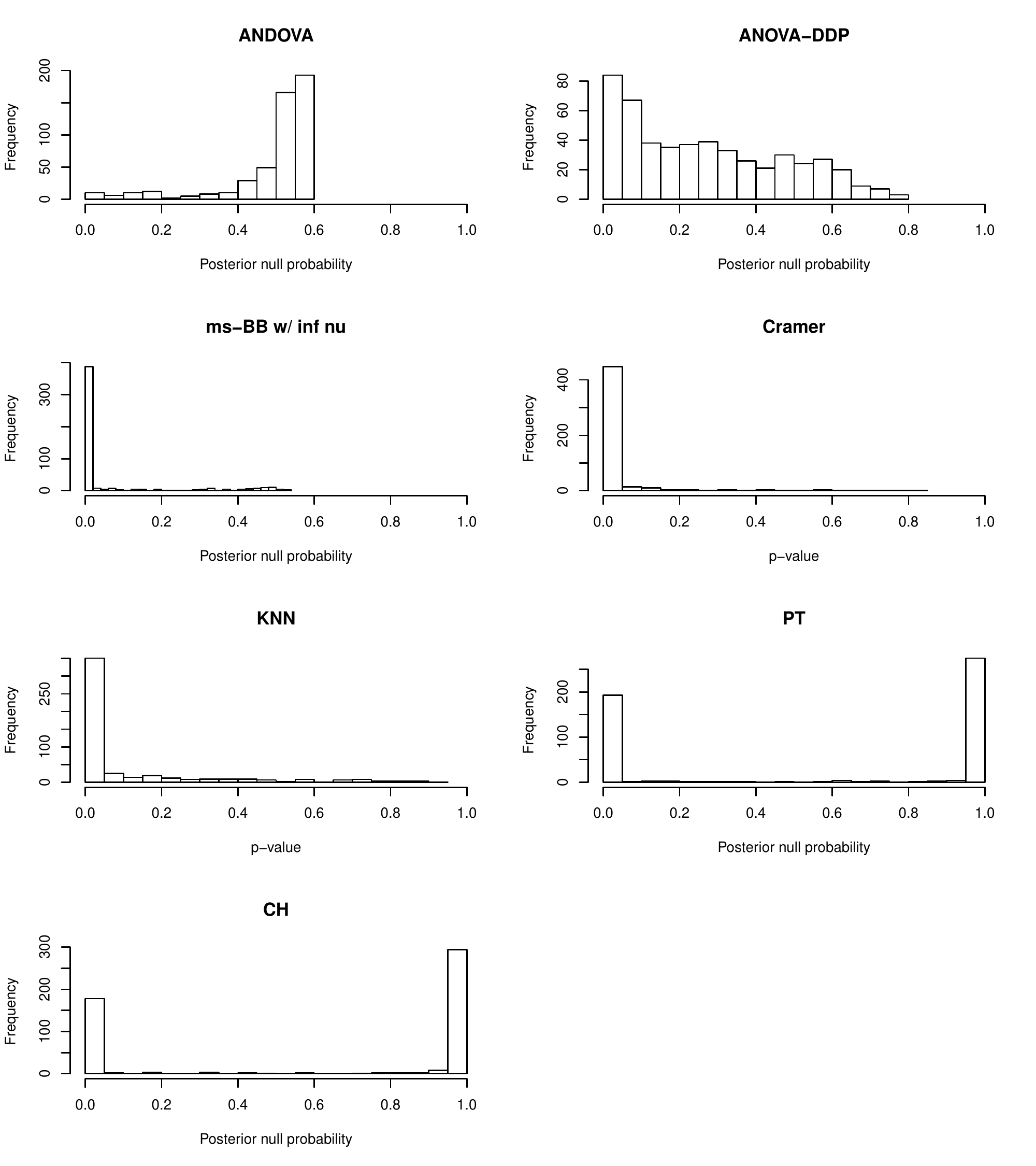}
\caption{The histogram of seven test statistics under the null hypothesis of no two-group difference. The seven methods are ANDOVA based on graphical ms-BB, ANOVA-DDP, ms-BB with $\nu(A)=\infty$), Cramer, KNN, PT, and CH.}
\label{fig:null_hist}
\end{center}
\end{figure}

\ref{fig:null_hist} presents the histogram of the corresponding test statistics of the different methods for testing the null hypothesis of no two-group difference when the null is indeed true. For our ANDOVA method as well as its $k$-sample counterpart with $\nu(A)=\infty$, the test statistic is the posterior joint probability of the null, i.e., 1$-$PJAP, and for the two frequentist tests Cramer and KNN, the corresponding $p$-value. The ANOVA-DDP is designed for estimation and prediction and under this model the factor effects is exactly zero with probability 0. As such, we define the null under this model to be the event that the factor effect is small in scale, namely less than 0.05 in absolute value, and we use the posterior probability of this event as the test statistic. Only ANDOVA behaves the way it should. The $k$-sample methods, which do not take into account the in-group variation,
display a high concentration of the test statistic near zero, showing strong {\em false} evidence against the null hypothesis. This is exactly the confounding phenomenon frequently occur in practice. We believe this is an important cause of irreplicable finding in cross-sample comparison in the scientific literature. By allowing in-group variation, ANDOVA effectively address this problem. The PJAP is pushed above the prior joint null probability of 50\% and to about 60\% in this case. (With more replicates, it will be pushed further toward 1.) ANOVA-DDP addresses the problem to some extent but still has a concentration near zero somewhat less extreme than the $k$-sample methods. 

\vspace{-1em}

\subsection{Testing performance for cross-group differences}
\label{sec:testing_performance}
\vspace{-0.5em}

In addition to ensuring the validity of inference under the null, ANDOVA can also improve the discriminatory ability in detecting actual cross-group differences. To examine this, we simulate data from four alternative scenarios, representing four common types of distributional differences---namely, (i) local shift, (ii) local dispersion difference, (iii) global shift, and (iv) global dispersion difference. In all scenarios, we again simulate two groups of data, with each group consisting of four replicate samples with in-group variation.

Under each of these four scenarios, the ``centroid'' distribution of Group~1 is still the mixture model given in \eqref{eq:null_centroid}, while that of Group~2 is given as follows (with bold font indicating the differing part from Group~1) 
\begin{itemize}
\item Scenario 1---Local shift:
\vspace{-1em}
\[
\frac{1}{3} \, {\rm N}({\bf 1.1},0.05^2) + \frac{1}{3}\, {\rm N}(1.5,0.2^2) + \frac{1}{3}\, {\rm N}(2.5,0.1^2).
\]
\vspace{-3.7em}

\item Scenario 2---Local dispersion difference:
\vspace{-1em}
\[
\frac{1}{3} \, {\rm N}(1,{\bf 0.15^2}) + \frac{1}{3}\, {\rm N}(1.5,0.2^2) + \frac{1}{3}\, {\rm N}(2.5,0.1^2).
\]
\vspace{-3.7em}

\item Scenario 3---Global shift:
\vspace{-1em}
\[
\frac{1}{3} \, {\rm N}({\bf 1.05},0.05^2) + \frac{1}{3}\, {\rm N}({\bf 1.55},0.2^2) + \frac{1}{3}\, {\rm N}({\bf 2.55},0.1^2).
\]
\vspace{-3.7em}

\item Scenario 4---Global dispersion difference:
\vspace{-1em}
\[
\frac{1}{3} \, {\rm N}(1,{\bf 0.1^2}) + \frac{1}{3}\, {\rm N}(1.5,{\bf 0.4^2}) + \frac{1}{3}\, {\rm N}(2.5,{\bf 0.2^2}).
\]
\end{itemize}
\vspace{-1em}

\begin{figure}[p]
\begin{center}
\includegraphics[width=33em]{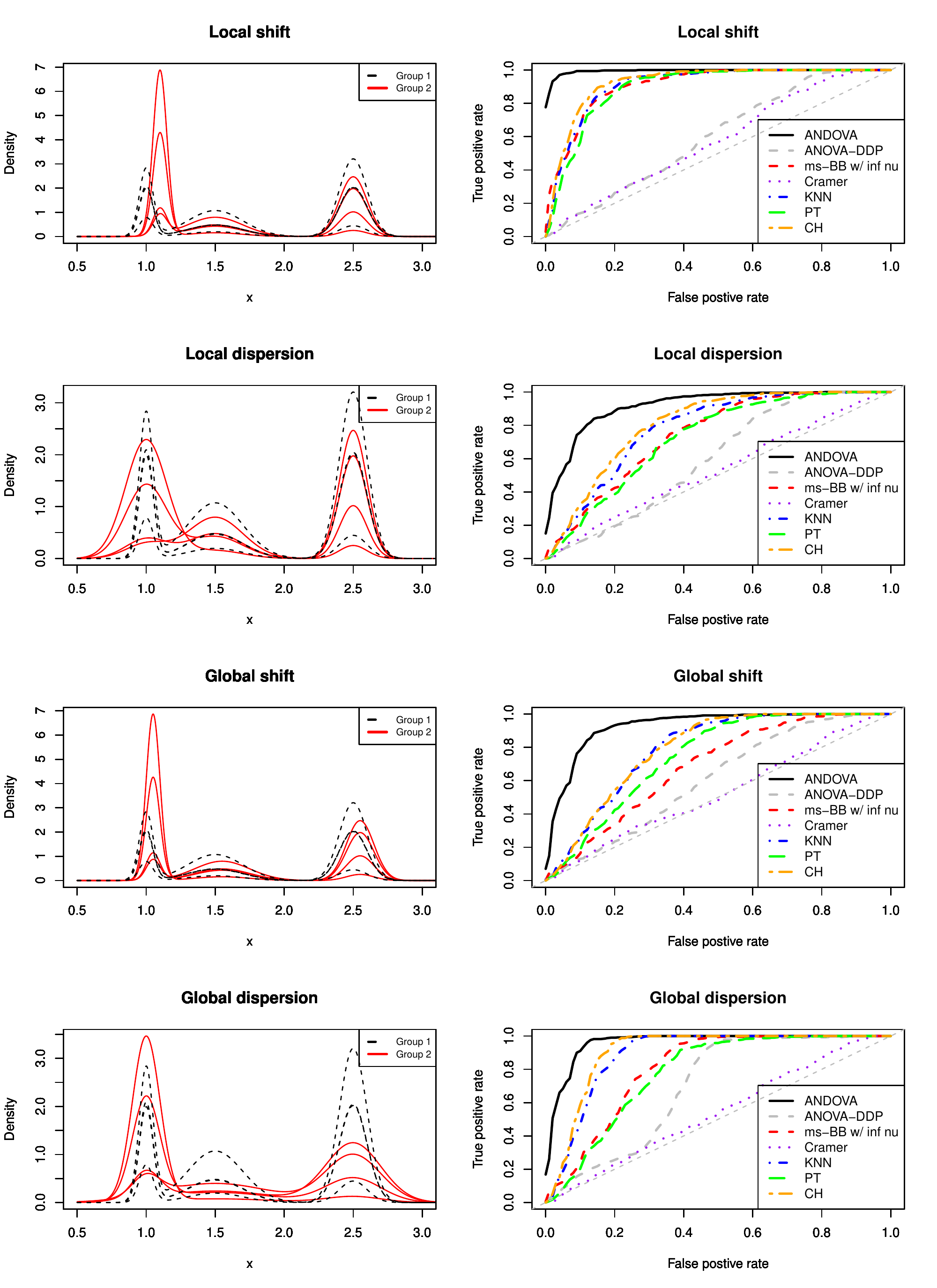}
\vspace{-0.5em}
\caption{Performance comparison of different methods under the four scenarios with cross-group differences. Left panel: A realization of the true sampling densities under the simulation setting for each alternative scenario. Group~1 densities are indicated with dashed black lines and Group~2 solid red lines. Each group consists of four replicate samples. Right: ROC curves of methods---ANDOVA based on graphical ms-BB, ANOVA-DDP, ms-BB with $\nu(A)=\infty$, Cramer, KNN, PT and CH---for detecting the existence of cross-group differences.}
\label{fig:roc_curves}
\end{center}
\end{figure}

\noindent We introduce in-group variation by adding random perturbations to the mixing weights in the same way as in Section~\ref{sec:null_behavior}. The left panel of \ref{fig:roc_curves} shows a typical example of sampling distributions drawn from the above simulation setting under each scenario.

We again generate 500 total observations split among the four replicate samples in each group, repeat the simulation under each scenario 500 times, and apply the seven methods in each simulation. We present ROC curves of the four test statistics in \ref{fig:roc_curves} constructed based on their values under each of the four alternative scenarios and those under the null scenario computed in the simulations in Section~\ref{sec:null_behavior}.

We see from the ROC curves that ANDOVA substantially outperforms all other methods. Cramer test suffers particularly badly in the presence of in-group variation, essentially losing all of its discriminatory ability in separating the alternatives from the null.

\begin{figure}[p]
  \begin{center}
  
\mbox{
  \subfigure[Local shift - ANDOVA]{\includegraphics[width=0.45\textwidth, clip=TRUE, trim=0 10mm 0 20mm]{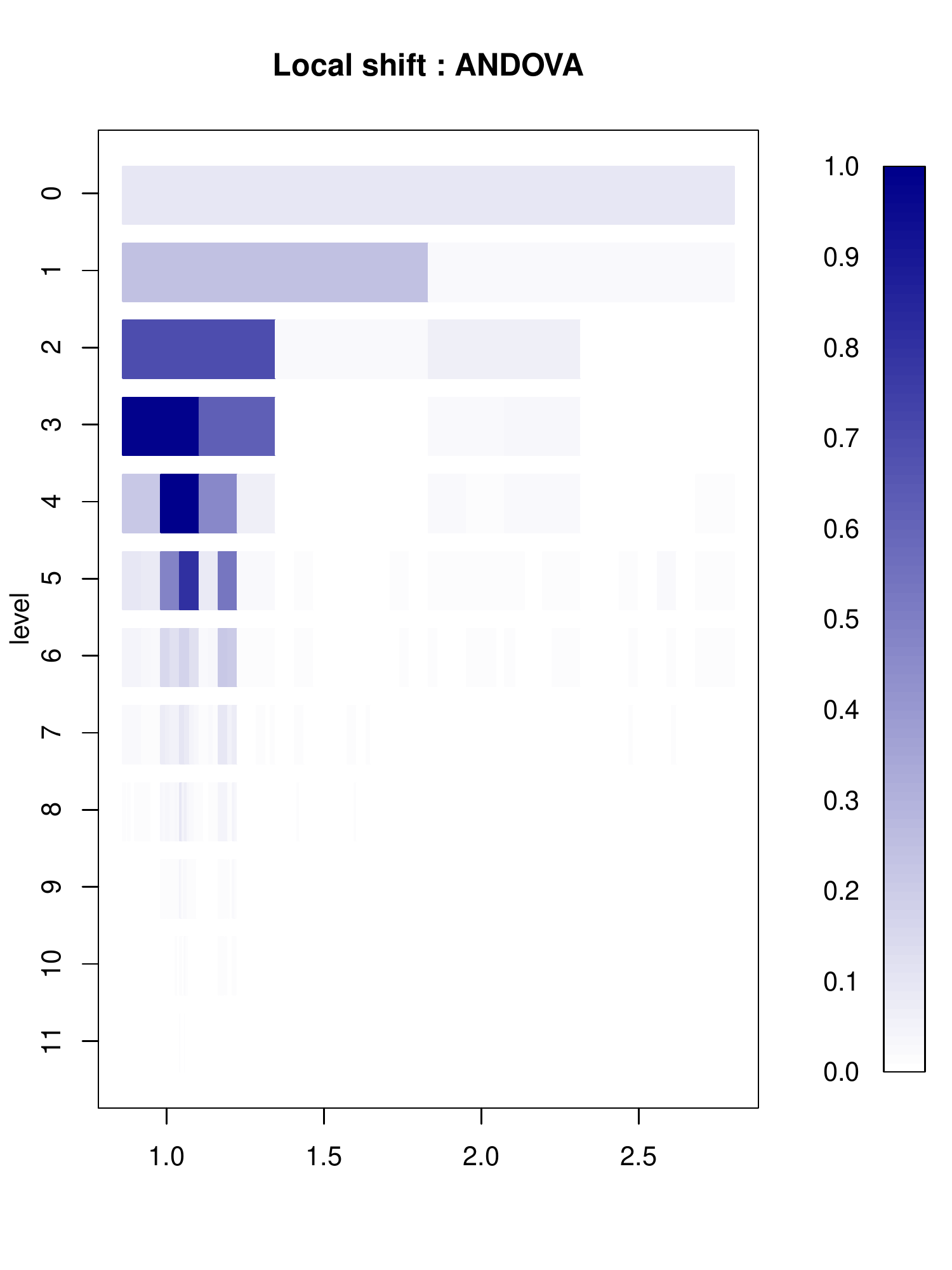}}
  \hspace{3em}
  \subfigure[Local shift - ms-BB with $\nu(A)=\infty$]{\includegraphics[width=0.45\textwidth, clip=TRUE, trim=0 10mm 0 20mm]{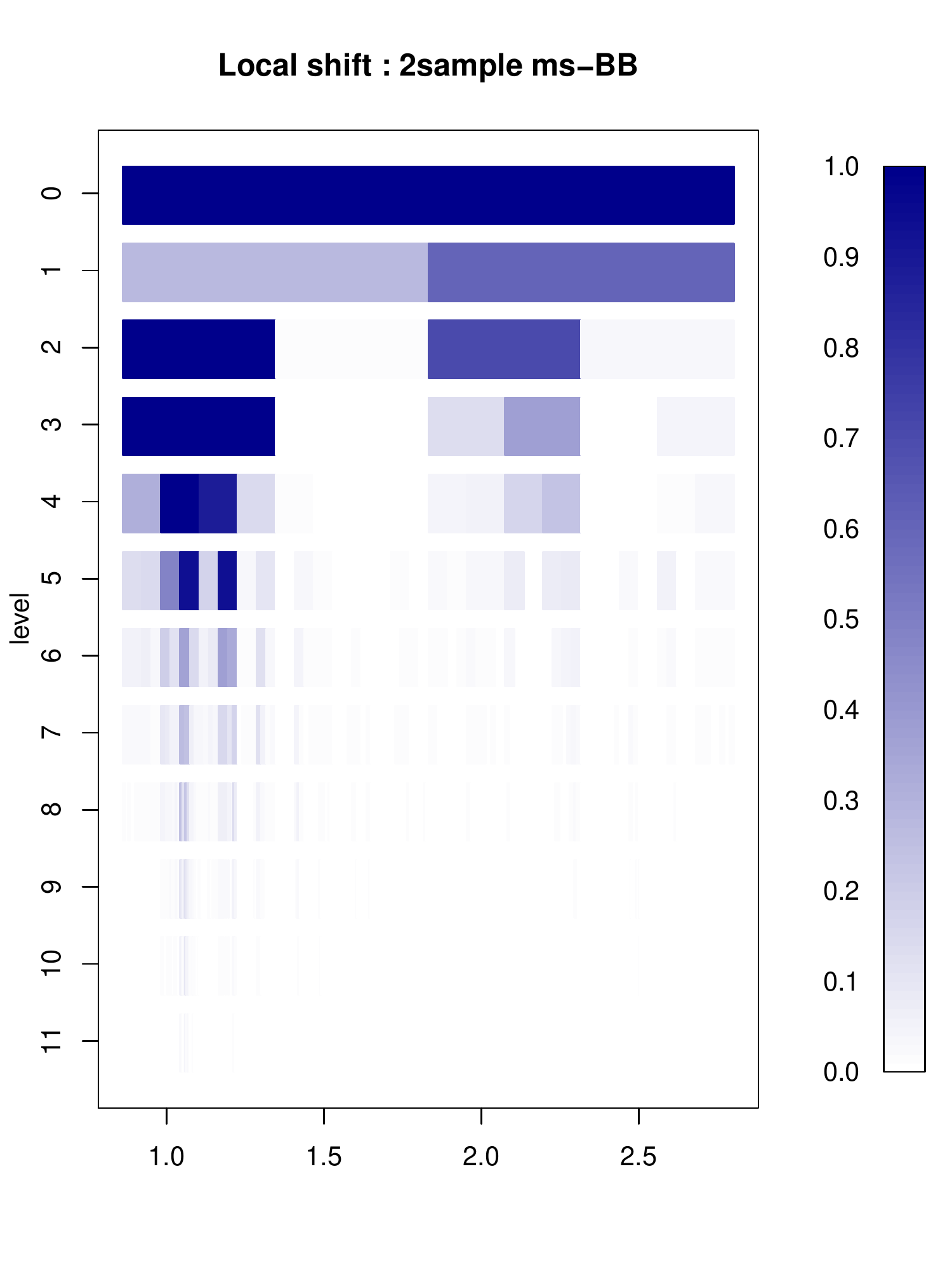}}
}
\vspace{1em}

\mbox{
  \subfigure[Local dispersion - ANDOVA]{\includegraphics[width=0.45\textwidth, clip=TRUE, trim=0 10mm 0 20mm]{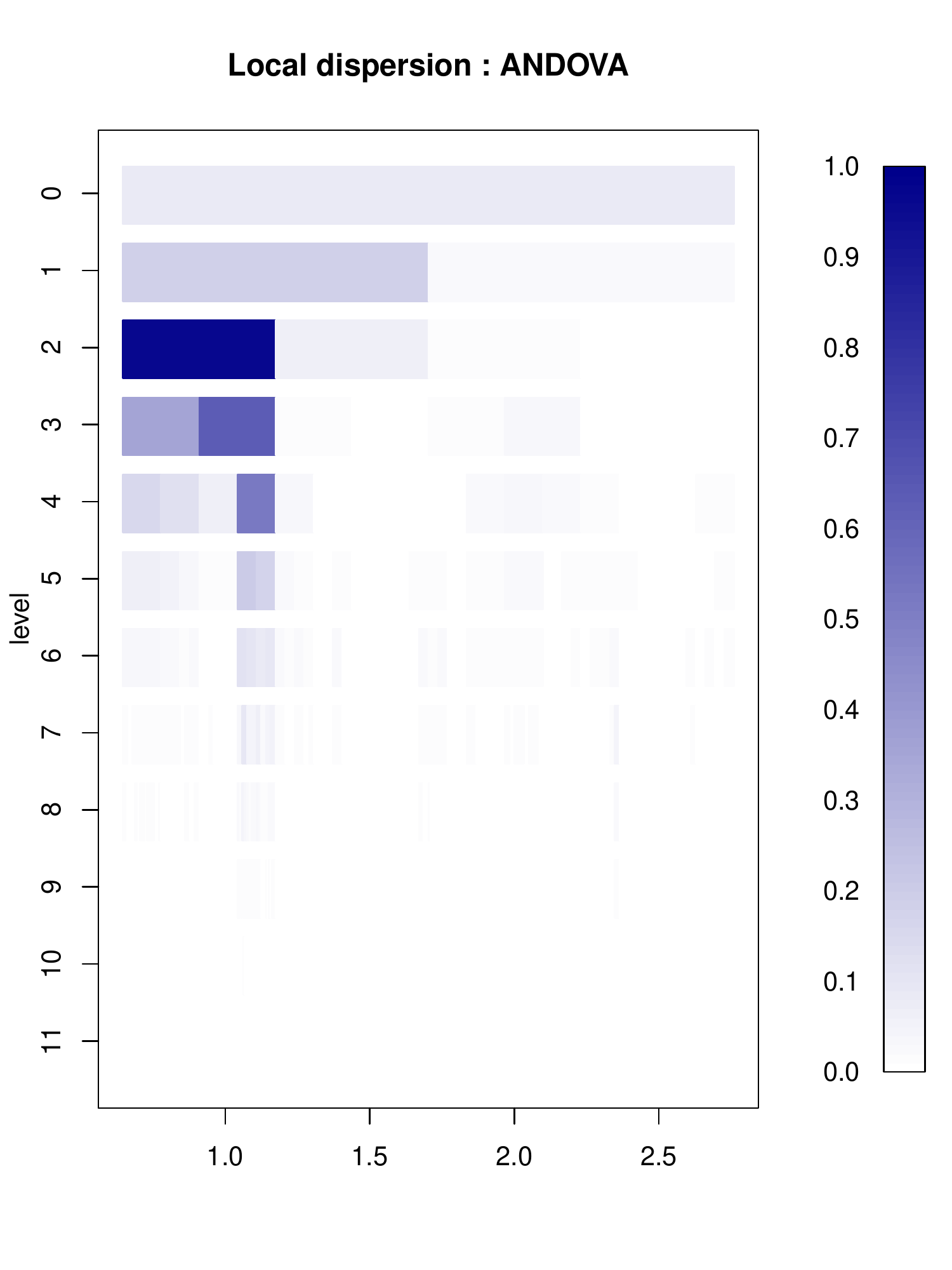}}
  \hspace{3em}
  \subfigure[Local dispersion - ms-BB with $\nu(A)=\infty$]{\includegraphics[width=0.45\textwidth, clip=TRUE, trim=0 10mm 0 20mm]{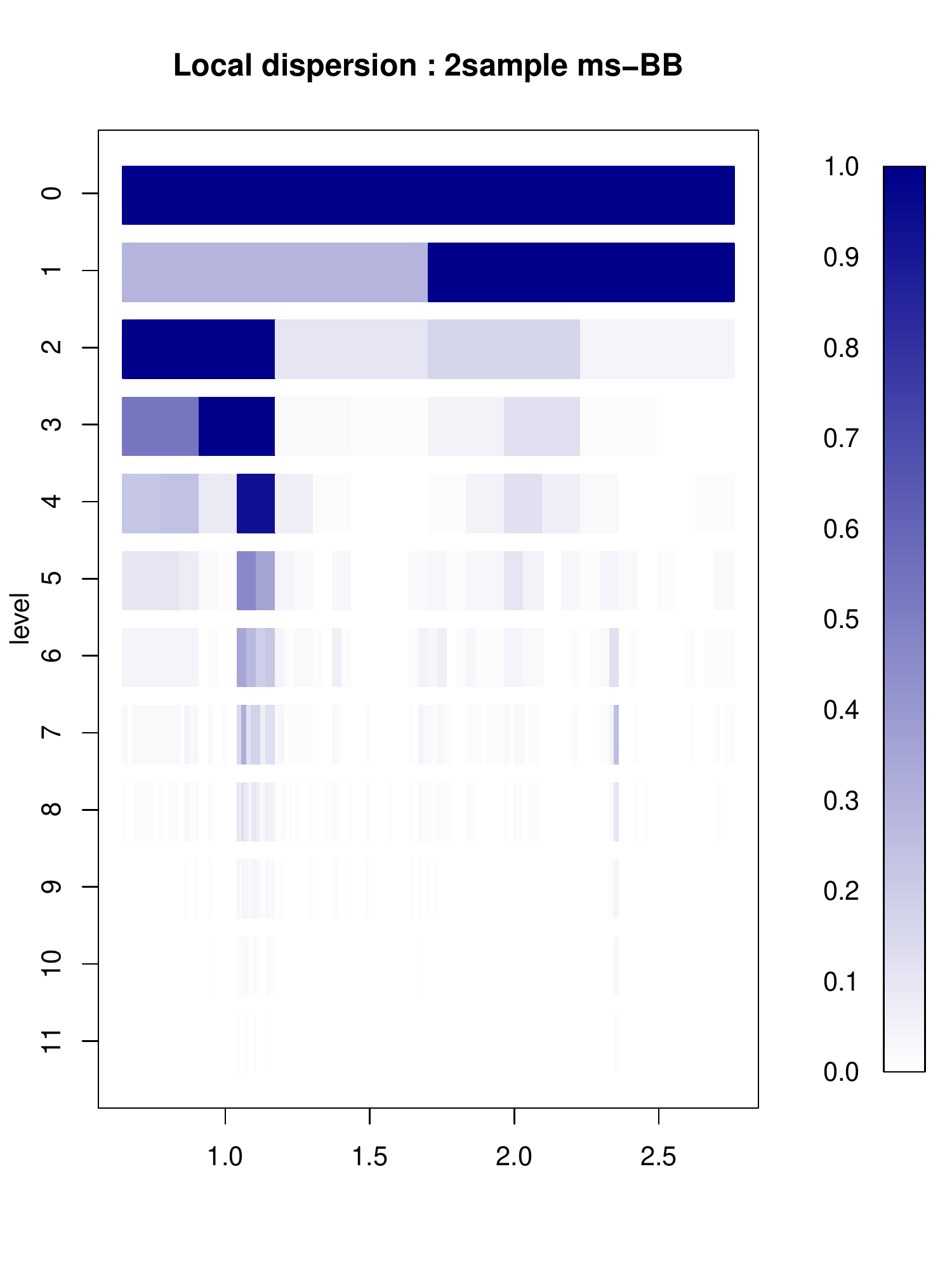}}

}
    \caption{PMAPs from ANDOVA and two-sample ms-BB under the two local alternative scenarios. Left column: ANDOVA under the graphical ms-BB model; right column: ms-BB with $\nu(A)=\infty$. Upper row: local shift; Lower row: local dispersion.}
    \label{fig:pmaps}
  \end{center}
\end{figure}

\vspace{-1em}

\subsection{Characterizing cross-group differences}
\label{sec:identifying_difference}
\vspace{-0.5em}

In many modern applications of cross-group comparisons, not only is it interesting to call for the existence of differences, but to identify the nature of the differences as well. This can be readily achieved under our graphical ms-BB framework for ANDOVA. Specifically, after computing the PMAP on each window, we can visually represent the posterior information regarding the cross-group difference by plotting the PMAPs on all of the windows up to some resolution on a multi-scale tree plot.

For instance, in \ref{fig:pmaps}(a) and (c) we plot the PMAPs from ANDOVA for the windows in a typical simulation run under the two local difference scenarios. In both cases, the PMAPs take large values in windows that indeed reflect the difference at the group level. In \ref{fig:pmaps}(b) and (d) we show the PMAPs for the 2-sample version of ms-BB with $\nu(A)=\infty$, In contrast, several windows with the highest PMAPs are in fact reflecting in-group variation and thus are false positives. This shows the importance of properly adjusting for in-group variation in order to identify the true structure of cross-group difference.

\vspace{-1em}

\section{Data analysis: Comparing DNase-seq count profiles}
\label{sec:dnase}
\vspace{-0.5em}

We next analyze a data set from the ENCODE project \citep{encode:2012} to study gene regulation in cancer cells. In particular, we compare three cell lines, two of which are from patients with brain tumors, while the other from chronic myelogenous leukemia. The nerve growth factor gene VGF is known to be activated (or expressed) in the two brain tumor cell lines, and suppressed in the other cell line. It is thus of interest in studying what and how transcriptional factors (TFs)---genes that control other genes' expression levels---differentially regulate the expression of VGF in these cell lines. 

DNase-seq \citep{song:2010,degner:2012} is a next-generation sequencing experimental protocol that uses sequencing counts to measure the whether chromatins are open as an indicator for TF binding events. Depending on the nature of the binding event, the sequencing counts will display different distributions at and near potential binding sites along the genome. Therefore, one strategy to investigate whether some TFs regulate a particular gene, e.g., VGF, is to compare the distribution of sequencing count distributions in regions with potential binding sites. 

\begin{figure}[p]
\begin{center}
  \includegraphics[width=38em]{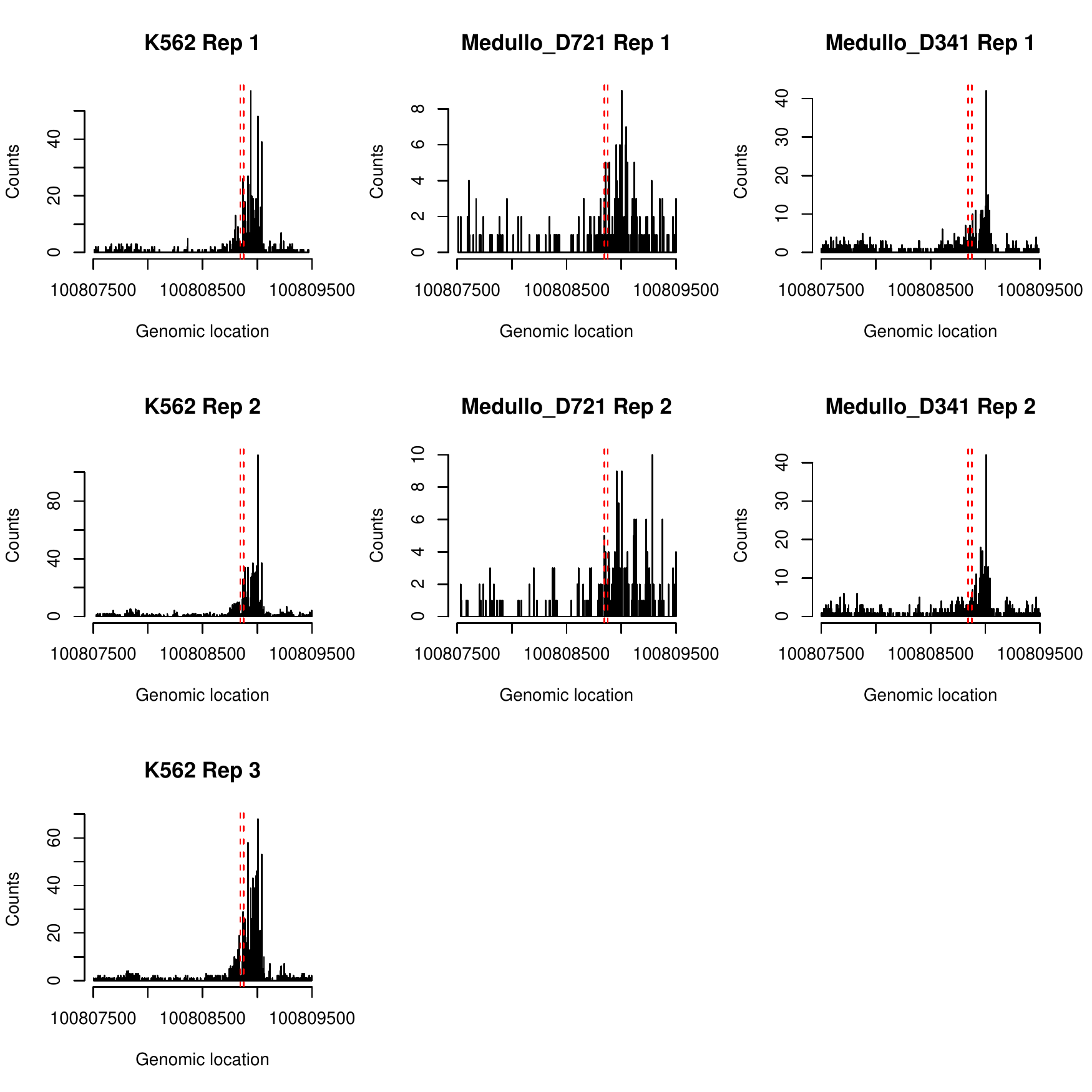}
\caption{The DNase-seq count histograms in the genomic region around gene VGF on human chromosome 7 for three cell lines. Each column corresponds to a cell line---K562 (leukemia), Medullo D721 (brain tumor 1), Medullo D341 (brain tumor 2). Each row corresponds to a replicate---K562 has three replicates, while the other two cell lines each have two. The red dashed lines mark the boundaries of the scanning window in level six called to be significant in cross-group difference in \ref{fig:dnase_counts}(a).}
\label{fig:dnase_counts}
\end{center}
\end{figure}

The data being analyzed here are the sequencing counts in the genomic region around the VGF gene on Chromosome 7. In particular, for the leukemia cell line (K562) three replicate samples are obtained while for each of the two brain tumor cell lines two replicate samples are available. \ref{fig:dnase_counts} presents the count distribution of the seven samples. We see that there is substantial within-group variation whose extent varies across the genomic locations.

\begin{figure}[h]
  \begin{center}
    \mbox{
      \subfigure[PMAPs from ANDOVA]{\includegraphics[width=0.42\textwidth,height=0.4\textwidth,clip=TRUE,trim=0mm 0mm 0mm 20mm]{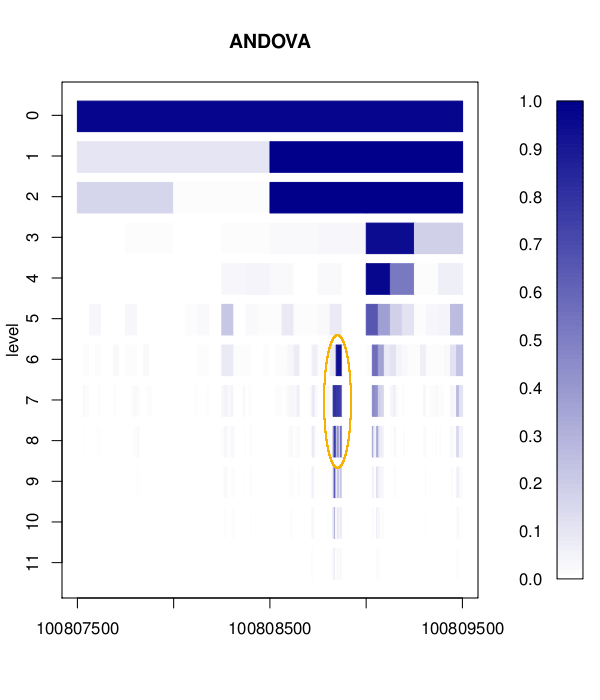}}\hspace{3em}
      \subfigure[PMAPs from ms-BB (with $\nu(A)=\infty$)]{\includegraphics[width=0.42\textwidth,height=0.4\textwidth,clip=TRUE,trim=0mm 0mm 0mm 20mm]{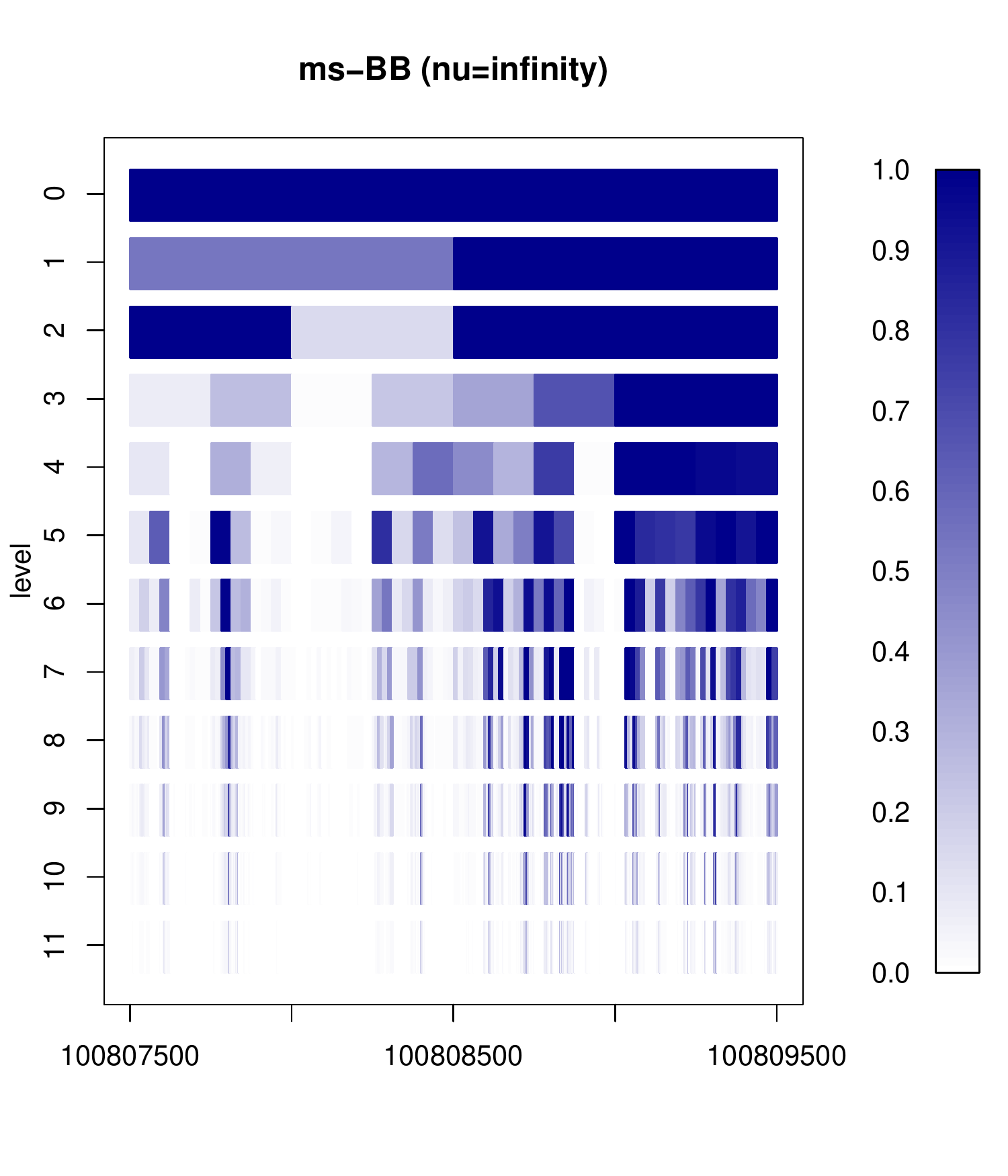}}
   }

  \caption{DNase example: (a) The PMAPs for ANDOVA on comparing the DNAse-seq count distribution near gene VGF on Chromosome 7 for three tumor cell lines---K562 (3 replicates), Medullo D721 (2 replicates), and Medullo D341 (2 replicates). The significant windows in levels 0, 1, and 2 capture the large scale difference in the distribution on the entire VGF gene as a whole. The string of significant windows in levels 6 to 8 (marked with a gold ellipse) correspond to a binding site of the NRSF repressor factor and is the same region marked by red dashed lines in \ref{fig:dnase_counts}. (b) The PMAPs for the 3-sample ms-BB with $\nu(A)=\infty$, incurring pervasive confounding due to ignoring in-group variation.}
\label{fig:dnase_pmaps}
  \end{center}
\end{figure}

\begin{figure}[!h]
  \begin{center}
    \mbox{
     \subfigure[K562]{\includegraphics[width=0.31\textwidth,clip=TRUE,trim=0mm 10mm 0mm 20mm]{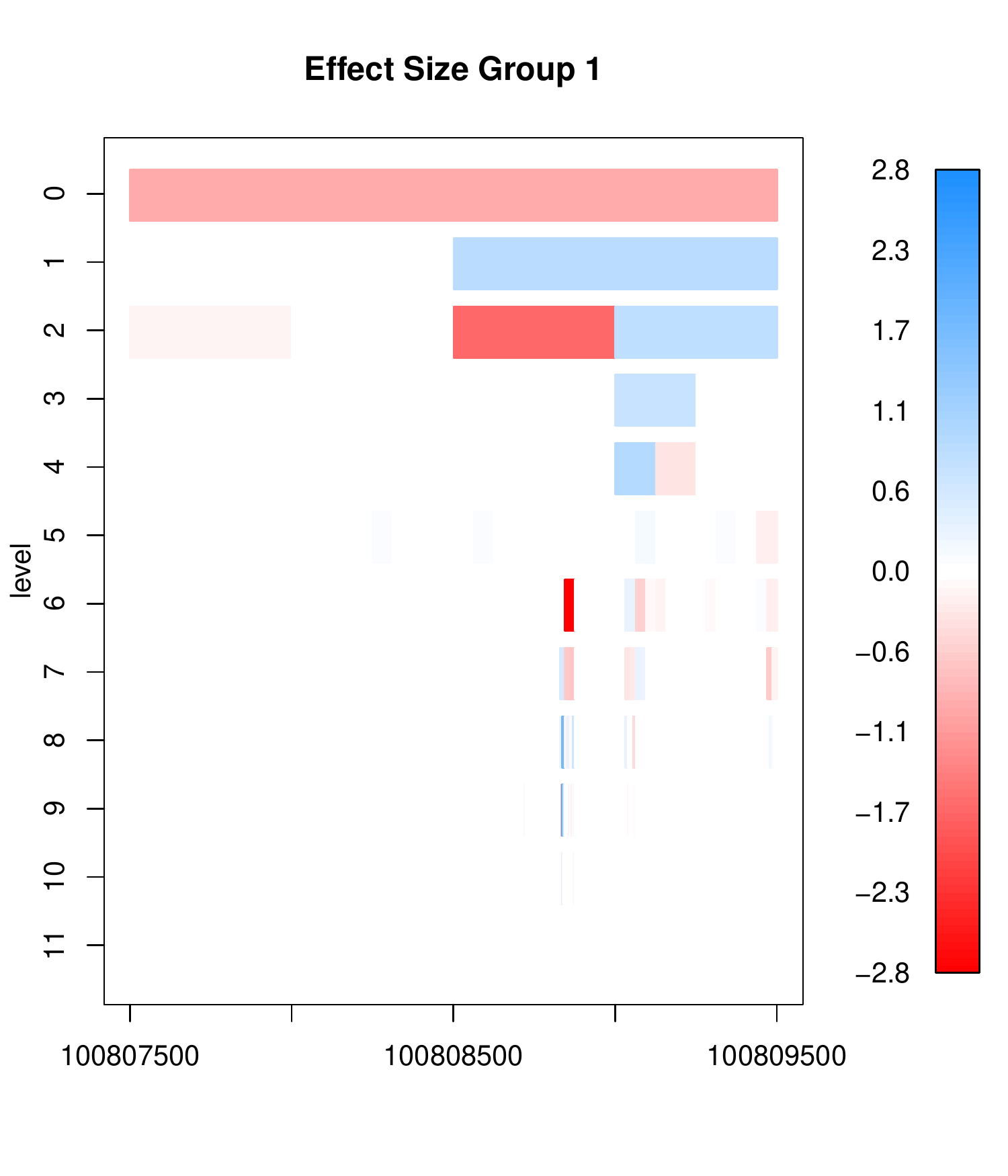}}\hspace{0.5em}
     \subfigure[Medullo D721]{\includegraphics[width=0.31\textwidth,clip=TRUE,trim=0mm 10mm 0mm 20mm]{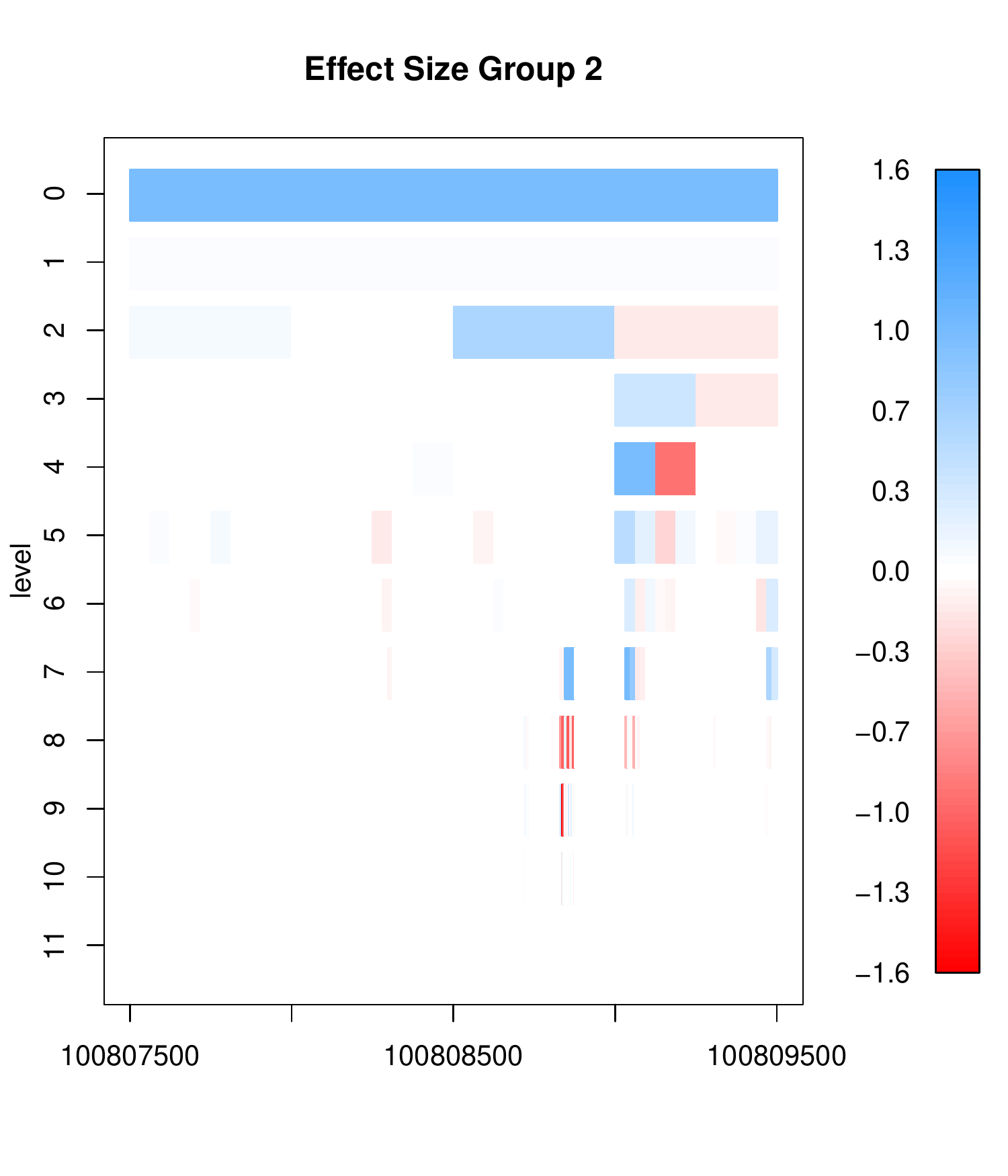}}\hspace{0.5em} 
     \subfigure[Medullo D341]{\includegraphics[width=0.31\textwidth,clip=TRUE,trim=0mm 10mm 0mm 20mm]{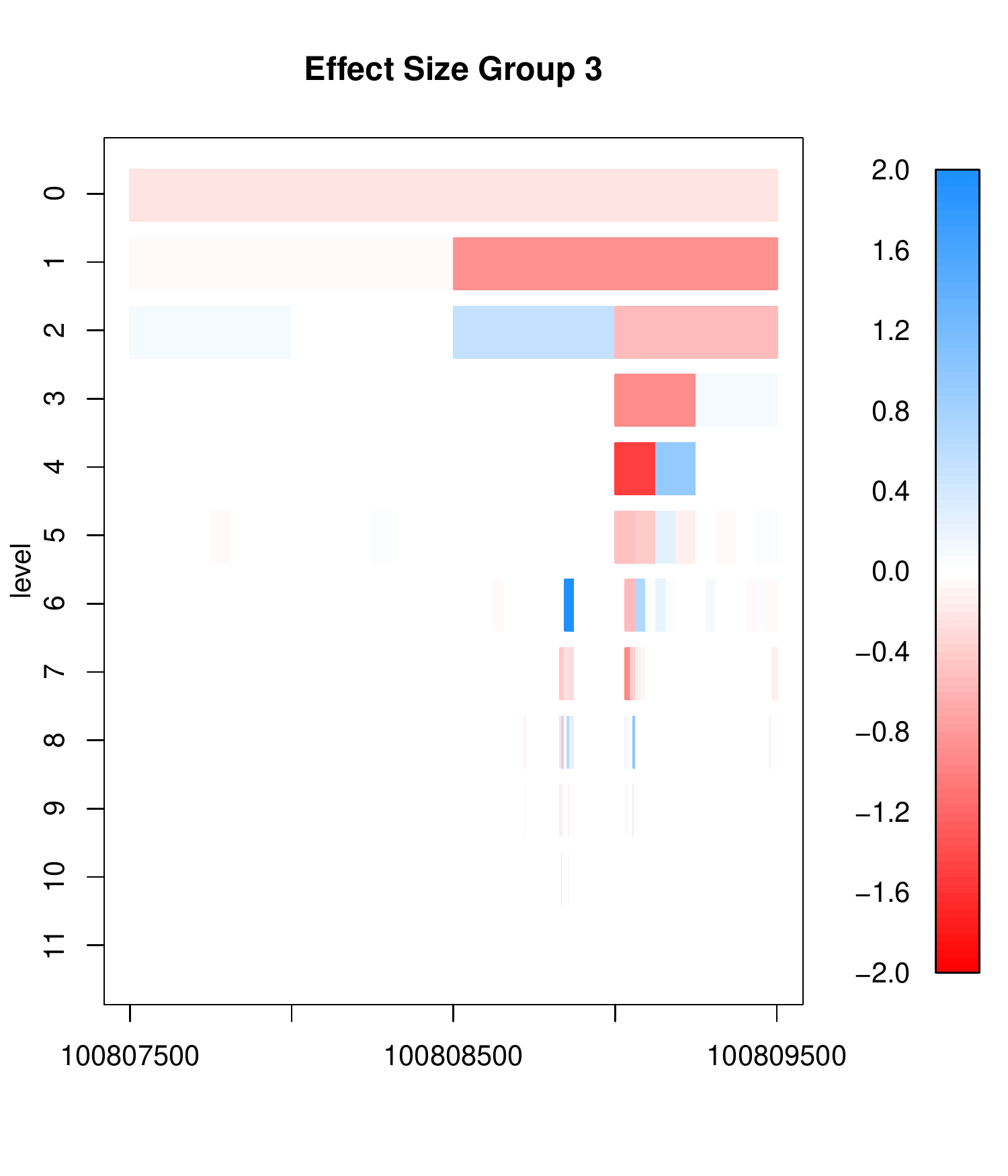}}
   }
\vspace{-0.5em}

  \caption{ANDOVA effect sizes for the three cell lines.}
\label{fig:dnase_eff_sizes}
  \end{center}
\end{figure}

This is in essence a 3-group ANDOVA problem. Our goal is to identify differences in the underlying count distributions across the groups. A tree-plot of the PMAPs from the ANDOVA under the graphical ms-BB are presented in \ref{fig:dnase_pmaps}(a). The tree plots for the effect sizes as evaluated in Eq.~\eqref{eq:post_effect_size} are given in \ref{fig:dnase_eff_sizes}. 
The string of significant windows in levels 6 to 8 (marked with a gold ellipse in \ref{fig:dnase_pmaps}(a)) are particularly interesting, as they shed light on what TFs may be binding in regulating the expression of VGF. In particular, that window covers a genomic location from 100,808,844 to 100,808,876, which overlaps exactly with a binding site in the promoter region of VGF for a TF called NRSF. NRSF is a repressor TF, which inhibits gene expression. 
The finding suggests that the NRSF is an important player in regulating VGF in neuronal vs non-neuronal cells. 

To illustrate the importance of accounting for in-group variation, we repeat the analysis while setting $\nu(A)=\infty$ for all windows, i.e., we apply the 3-sample version of the ms-BB model. This model ignores the in-group variation and combines the observations from all samples in each group into a single sample representing that group. The corresponding PMAPs are presented in \ref{fig:dnase_pmaps}(b). There are a large number of ``significant'' windows that are false positives due to the confounding of in-group variation with cross-group difference.

\vspace{2em}

\vspace{-3em}

\section{Concluding remarks}
\vspace{-0.5em}

We have introduced a multi-scale modeling approach to addressing ANDOVA. Our method uses a decomposition of probability distributions to transform the general ANDOVA into Bayesian inference on the ms-BB model. Moreover, a graphical hyperprior is incorporated into the ms-BB model to achieve borrowing of strength across locations and scales. Using a combination of numerical integration and message passing, Bayesian inference under the graphical ms-BB model can be carried out efficiently. The computation required is linear in the number of total sample size, making it applicable to large-scale problems such as genomics where the comparison often needs to be completed at a large number of genomic locations.

The proposed framework can be generalized to multi-dimensional distributions. The main complication is the construction of windows on the sample space. To this end, previous works \citep{walther2010optimal,rufibach2010block} show that an effective strategy is through bisecting windows along each dimension of the sample space. One can incorporate an additional layer of modeling on the underlying partitioning into the hierarchical model, by adding a hyperprior on the partitioning, to allow inference on the proper windowing strategy \citep{wong_ma_2010}. Indeed, the numerical integration and message passing algorithms will still apply to provide computational recipe for evaluating the joint posterior and in particular the PMAPs.

\vspace{-1.5em}

 \section{Acknowledgment}
\vspace{-0.5em}

The authors are grateful to Kevin Luo for helping with extracting the DNase data set from the ENCODE project website and for providing the biological motivation for the analysis.
\vspace{-1.5em}

\begingroup
    \setlength{\bibsep}{2pt}
    \bibliography{andova}
\endgroup

\newpage

\setcounter{page}{1}

\subsection*{Supplementary Materials}
 \subsubsection*{S1.~Proofs}
 \begin{proof}[Proof of Theorem~1]
Our proof shall proceed in two steps. First we prove the result for a finite NDP $\T$ with a maximum resolution level $K$. Then we extend the result to infinite NDPs. Suppose $\T$ is a finite NDP with maximum resolution $K$. For each $A\in\T$ and $s\in\{0,1\}$, let 
\vspace{-2em}

\begin{align*}
\phi_{s}(A) := {\rm Pr}(\bx(A)\,|\,A,S(A_p)=s)\Big/\prod_{A\in\T(A)} M_0(A)
\end{align*}
\vspace{-2em}

\noindent where $\T(A)$ denotes the subtree of $\T$ consisting of $A$ (as the root) and all its descendants. 
When $\sum_{j} n_{ij}(A)>0$ for no more than one $i$, it is easy to check that $M_0(\tilde{A})=M_1(\tilde{A})$ for all $\tilde{A}\in\T(A)$, and so 
\[
{\rm E}\left(\prod_{A\in\T^{(k)}(A)} M_{S(A)}(A)\,|\,S(A_p)=s\right) = \prod_{A\in\T^{(k)}(A)} M_0(A) \quad \text{for all $k$.}
\]
Hence in that case $\phi_0(A)=\phi_1(A)=1$.

When $\sum_{j} n_{ij}(A)>0$ for at least two $i$'s, and if $A$ has children $A_l$ and $A_r$, we have
\vspace{-3em}

\[
{\rm Pr}(\bx(A)\,|\,A,S(A_p)=s) = \sum_{s'}\rho_{s,s'}(A) M_{s'}(A) {\rm Pr}(\bx(A_l)\,|\,A,S(A)=s')\cdot {\rm Pr}(\bx(A_r)\,|\,A,S(A)=s').
\]
\vspace{-3em}

\noindent Thus 
\vspace{-3.5em}

\begin{align*}
\phi_s(A) &= \sum_{s'}\rho_{s,s'}(A) \cdot \frac{M_{s'}(A)}{M_0(A)} \cdot \frac{{\rm Pr}(\bx(A_l)\,|\,A,S(A)=s')}{\prod_{A\in\T(A_l)} M_0(A_l)}\cdot \frac{{\rm Pr}(\bx(A_r)\,|\,A,S(A)=s')}{\prod_{A\in\T(A_r)} M_0(A_r)}\\
&= \rho_{s,0}(A)\phi_0(A_l)\phi_0(A_r) + \rho_{s,1}(A)\cdot {\rm BF}(A) \cdot \phi_1(A_l)\phi_1(A_r).
\end{align*}
\vspace{-3em}

\noindent When $A$ has no children, i.e., $A\in\T^{K}$, the above equality holds with $\phi_s(A_l)=\phi_s(A_r)=1$.

By Bayes theorem,
\vspace{-1em}
\begin{align*}
P(S(A)=s'|S(A_p)=s,\bx)&=P(S(A)=s',\bx(A)\,|\,A,S(A_p)=s)/P(\bx(A)\,|\,A,S(A_p)=s)\\
&=\rho_{s,s'}(A) \cdot \frac{M_{s'}(A)}{M_0(A)}\cdot \phi_{s'}(A_l)\phi_{s'}(A_r)/\phi_s(A).
\end{align*}
\vspace{-3em}

\noindent Letting $\bphi(A)= (\phi_{0},\phi_{1})'$ and putting the above equality in matrix form completes the proof for the case when $\T$ is finite with a maximum resolution $K$.

Now consider the case when $\T$ does not terminate at a finite maximum resolution $K$. Then we use a limiting truncation argument as follows. Now for each $k=1,2,\ldots$, and each $A\in\T^{(k)}$
\vspace{-4em}

\begin{align*}
\phi_{s}^{(k)}(A) := {\rm Pr}(\bx(A)\,|\,A,S(A_p)=s,\T=\T^{(k)})\Big/\prod_{A\in\T^{(k)}(A)} M_0(A)
\end{align*}
\vspace{-3em}

\noindent where $\T^{(k)}(A)$ is the subtree of $\T^{(k)}$ rooted at $A$. Now define
\vspace{-1em}
\[
\phi_{s}^{(\infty)}(A) = \lim_{k\rightarrow \infty} \phi_{s}^{(k)} = \lim_{k\rightarrow \infty} {\rm Pr}(\bx(A)\,|\,A,S(A_p)=s,\T=\T^{(k)})\Big/\prod_{A\in\T^{(k)}(A)} M_0(A).
\]
\vspace{-3em}

\noindent Because when $\sum_{j} n_{ij}(A)>0$ for no more than one $i$, $M_0(\tilde{A})=M_1(\tilde{A})$ for all $\tilde{A}\in \T(A)$, in that case we have $\phi_{s}^{(\infty)} = 1$. Now for $\sum_{j} n_{ij}(A)>0$ for at least two $i$'s,
\vspace{-3em}

\begin{align*}
&\phi_{s}^{(\infty)}(A)\\
=& \lim_{k} \sum_{s'}\rho_{s,s'}(A) \cdot \frac{M_{s'}(A)}{M_0(A)} \cdot \frac{{\rm Pr}(\bx(A_l)\,|\,A,S(A)=s',\T=\T^{(k)})}{\prod_{A\in\T^{(k)}(A_l)} M_0(A_l)}\cdot \frac{{\rm Pr}(\bx(A_r)\,|\,A,S(A)=s',\T=\T^{(k)})}{\prod_{A\in\T^{(k)}(A_r)} M_0(A_r)} \\
=& \rho_{s,0}(A)\phi^{(\infty)}_0(A_l)\phi^{(\infty)}_0(A_r) + \rho_{s,1}(A)\cdot {\rm BF}(A) \cdot \phi^{(\infty)}_1(A_l)\phi^{(\infty)}_1(A_r)
\end{align*}
where the last equality follows from moving the limit into the sum. Letting $\bphi(A)= (\phi_{0}^{(\infty)},\phi_{1}^{(\infty)})'$ completes the proof for the case with infinite $\T$.
\end{proof}
 

\end{document}